# PerovAgent: Autonomous Ligand Design for Halide Perovskite Nanocrystals

Seungjun Cha,[1,2] Yoo Jae Jeong,[3,4] Amanda Conde Del Moral,[1] Juan-Pablo Correa-Baena,[1,5] Seth R. Marder,[3,4,6,7] David S. Ginger,[8,*] Victor Fung,[2,*] Guoxiang Hu[1,5,*]

[1] School of Materials Science and Engineering, Georgia Institute of Technology, Atlanta, GA 30332, United States
[2] School of Computational Science and Engineering, Georgia Institute of Technology, Atlanta, GA 30332, United States
[3] Materials Science and Engineering Program, University of Colorado Boulder, Boulder, CO 80309, United States
[4] RASEI; The CU-Boulder/NLR Energy Institute, University of Colorado Boulder, Boulder, CO 80309, United States
[5] School of Chemistry and Biochemistry, Georgia Institute of Technology, Atlanta, GA 30332, United States
[6] Department of Chemistry, University of Colorado Boulder, Boulder, CO 80309, United States
[7] Department of Chemical and Biological Engineering, University of Colorado Boulder, Boulder, CO 80309, United States
[8] Department of Chemistry, University of Washington, Seattle, WA 98195, United States

* *Corresponding authors*
Email: dginger@uw.edu, victorfung@gatech.edu, emma.hu@mse.gatech.edu

**Abstract**

Designing surface-capping ligands for high-performance halide perovskite nanocrystals requires navigating a vast molecular space while discovering design rules that govern ligand-surface interactions. Here, we introduce PerovAgent, an autonomous multi-agent framework that combines large language model (LLM)-driven scientific reasoning with computational tools for atomistic model construction, machine-learned interatomic potential evaluation, and statistical validation. Across ten independent campaigns on $CsPbBr_3$ surfaces, PerovAgent adaptively explored and refined diverse ligand chemistries while progressively improving the binding-energy objective. Beyond optimization, the framework converts recurring chemical observations into explicit design principles that can be statistically evaluated. Of 35 formulated principles, 19 were prospectively supported, of which seven also passed the confound audit, highlighting the need to systematically validate plausible LLM-generated hypotheses during autonomous discovery. A recurrent alkynyl-phosphonate principle was further examined using density functional theory, which reproduced the predicted energetic preference. These results demonstrate an autonomous discovery framework that not only searches molecular space and optimizes the scalar objective but also transforms its search trajectory into systematically testable design principles. The framework can also be extended to more complex ligands, including zwitterionic, multidentate, and polymerizable ligands, to explore a wider range of ligand–surface interactions and uncover new design principles for halide perovskite nanocrystals.

## Introduction

Halide perovskite nanocrystals combine attractive optoelectronic properties, including bright emission, narrow linewidths, and high photoluminescence quantum yields, with facile solution processability.[1,2] Their colloidal synthesis typically relies on organic surface-capping ligands that provide colloidal stability in organic solvents[2,3], enabling low-cost production of highly monodisperse nanocrystals for applications in photovoltaics[4], light-emitting diodes[5], and quantum light sources.[6] Their practical performance, however, depends critically on the surface chemistry between the inorganic lattice and their ligand shell.[7] Labile ligands can be easily detached and create surface defects[3], particularly halide vacancies[8,9], introducing nonradiative recombination pathways and reactive sites for moisture and oxygen that accelerate degradation.[10] Surface ligands must therefore interact sufficiently strongly with the inorganic core to stabilize this interface through coordination and defect passivation.[11] Their role also extends beyond passivation, as ligand binding strength can influence properties such as interfacial charge distribution[11] and phonon dynamics of the inorganic lattice[12,13], which are closely related to the optoelectronic performance of halide perovskites. Designing effective ligands with suitable binding, and more importantly identifying the principles that govern ligand-surface interaction, has therefore become a central challenge in realizing high-performance halide perovskite optoelectronics.

Ligand design is inherently a discovery problem, fundamentally challenging because of the vast size of the feasible design space.[14,15] Exhaustive experimental screening is largely prohibitive, leaving the search heavily reliant on chemical knowledge and intuition of human experts, which makes the process costly and time-consuming. Data-driven approaches, particularly machine learning (ML), have substantially accelerated materials discovery by learning from high-throughput experimental measurements[16,17] or theoretical calculations using density functional theory (DFT), which is a well-established atomistic simulation technique.[18,19] More recently, advances in machine-learned interatomic potentials (MLIPs), which can achieve DFT-level accuracy at a fraction of computational cost, have enabled increasingly large numbers of candidate materials to be evaluated.[20] Nevertheless, the molecular design space remains far too large for brute-force exploration.[14,21] This challenge is particularly pronounced for ligand-passivated halide perovskites because ligand binding is not simply a property of an isolated molecule but involves interactions with the perovskite surface, which further expands the relevant configurational space and computational cost.[11,15] As computational evaluation becomes increasingly scalable and accurate, the primary bottleneck shifts from evaluation alone toward efficiently and adaptively navigating the design space: choosing where to explore, which candidates to propose, and how to refine the search based on the outcomes of previous evaluations.[21–23] These are precisely the decisions that scientists have traditionally made through iterative hypothesis formation, experimentation, and reasoning based on chemical domain knowledge.

Large language models (LLMs) offer a promising way to bring this scientific decision-making process into a computational materials discovery loop. Recent advances in LLMs, particularly in agentic systems, have demonstrated their ability to integrate broad domain knowledge, reason over intermediate results, formulate hypotheses, and coordinate external computational tools across various materials science problems.[24–28] This is particularly attractive for molecular discovery because organic molecules can be represented directly in text-based formats such as SMILES[29] and SMARTS[30], allowing LLMs to operate directly on molecular representations and perform targeted structural modifications within the same representation space.[28,31] Such an iterative reason-and-act (ReAct) loop[32] offers a route to efficiently navigate the otherwise intractable ligand design space through adaptive, chemically informed, and human-interpretable decision-making.

In this work, we introduce PerovAgent, a multi-agent framework that searches for strongly binding ligands while formulating structure–binding design principles for $CsPbBr_3$ halide perovskite surfaces. PerovAgent integrates three cooperating LLM agents that adaptively alternate between (1) exploring diverse binding groups and molecular backbones, (2) optimizing promising candidates through targeted structural modifications, and (3) testing emerging chemical hypotheses. Ligand binding energies and coordination geometries are evaluated using a pretrained MLIP with dispersion corrections, and the resulting feedback is incorporated into subsequent decisions through in-context learning. We introduce a two-stage statistical evaluation procedure that systematically tests LLM-generated hypotheses during autonomous discovery, with supported principles further informing subsequent molecular search. Across ten independent campaigns, PerovAgent identified several statistically supported design principles from recurring hypotheses, one of which was further examined with higher-fidelity DFT calculations. Together, we demonstrate that our framework enables adaptive exploration and refinement of molecular design space while explicitly deriving falsifiable, systematically testable, and interpretable ligand design principles for high-performing halide perovskite nanocrystals.

# Results and Discussion

## Multi-agent discovery framework

PerovAgent is designed to autonomously navigate molecular design space toward two coupled objectives: 1) optimizing the binding energy of surface ligands that passivate a halide vacancy on a $CsPbBr_3$ surface, and 2) deriving and testing structure–binding design principles from campaign outcomes. It operates as a closed-loop framework that integrates LLM-driven scientific reasoning with predefined computational tools (Figure 1). PerovAgent employs three LLM agents (Decider, Proposer, and Summarizer), each instantiated with distinct system prompts based on its intended role. These agents communicate through a shared state called the Blackboard, which stores campaign outcomes such as evaluated molecules, action and failure history, and the remaining evaluation budget. A full list of information stored in the Blackboard and available to each agent is summarized in Table S1. Throughout each campaign, the selection of subsequent actions and the formulation of chemical hypotheses are driven by LLM reasoning, whereas ligand validation, model construction, atomistic evaluation, statistical testing, and confound auditing are performed programmatically.

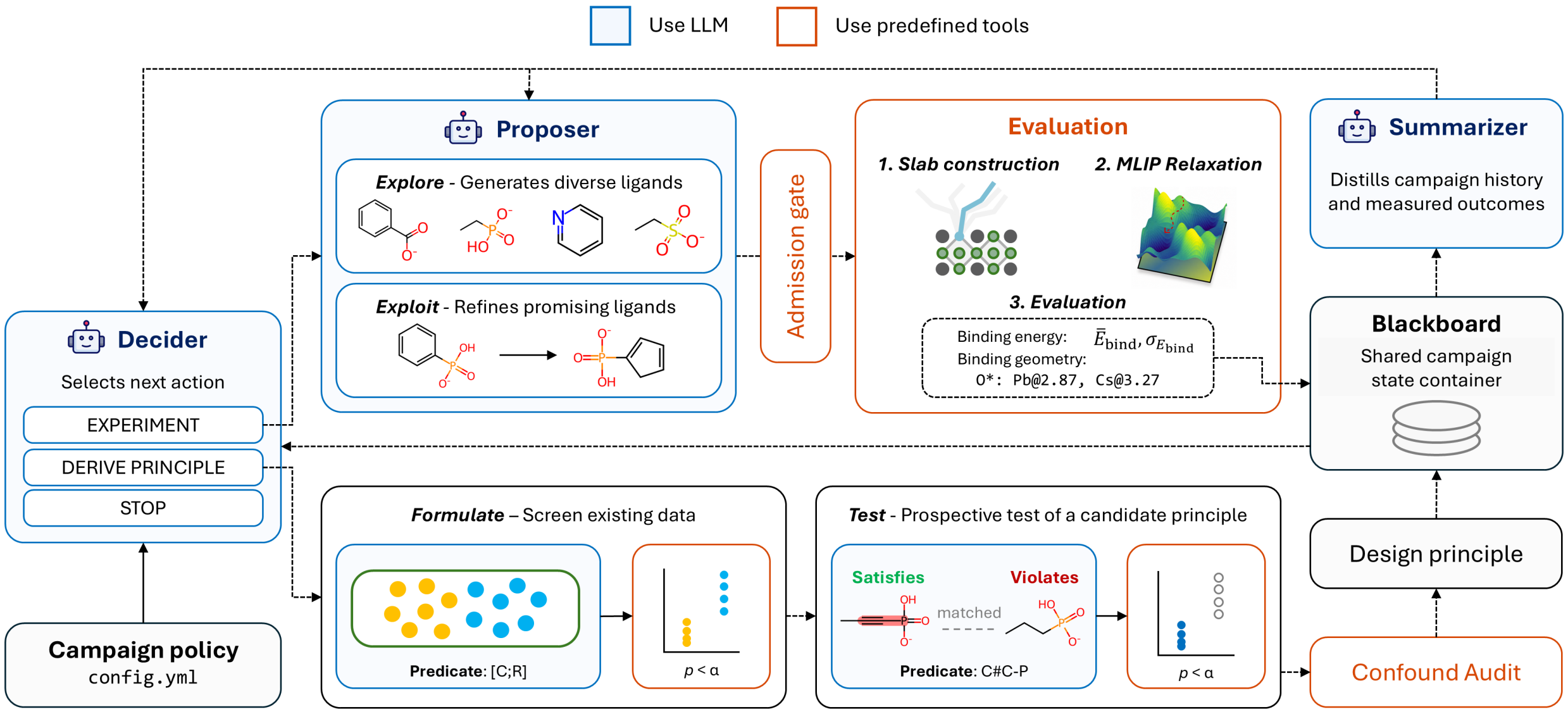

**Figure 1: Overview of the PerovAgent framework.** Given the input campaign policy, the Decider takes one ReAct step per iteration based on the current Blackboard state, choosing whether to run an experiment, derive a design principle, or stop the campaign. For an *Experiment*, the Decider provides a rationale to the Proposer, which generates candidate molecules in *Explore* or *Exploit* mode. Candidates pass an admission gate for validation before construction of a ligand-passivated $CsPbBr_3$ (001) slab, whose geometry is relaxed using a dispersion-corrected MLIP and subsequently evaluated for ligand binding energy and coordination geometry. For *Derive Principle*, a proposed principle, expressed as a structural predicate, is first tested during *Formulate* for statistical significance against the existing pool of evaluated molecules. Candidate principles are then subjected to prospective testing using newly designed matched molecular pairs. The Summarizer distills campaign outcomes and history into a compact playbook that is used to inform subsequent agent decisions.

Each run is initialized from an input configuration file that defines the campaign policy, including the LLM settings, campaign hyperparameters, ligand design space constraints, and evaluation budget (Table S2). At each iteration, the Decider reads the current Blackboard state and takes a single ReAct step, selecting one of three actions: run an *Experiment*, *Derive Principle* by formulating and testing a structural hypothesis, or *Stop* the campaign. This action space allows PerovAgent to adaptively choose the next course of action based on the current campaign context, rather than following a predetermined sequence. The campaign terminates when the Decider emits a *Stop* action, or when the evaluation budget is exhausted.

When an *Experiment* is selected, the Decider provides a rationale to the Proposer, which generates a ligand in one of two modes: *Explore* or *Exploit*. *Explore* broadens the accessible chemical space by proposing diverse binding groups and molecular backbones that are underrepresented in the current pool, whereas *Exploit* refines promising parent ligands through targeted structural modifications. Each proposal includes a SMILES representation of the ligand, its proposed binding anchor and binding motif, and a hypothesis regarding its expected performance compared to the parent ligand, if applicable. Importantly, ligand generation is not restricted to a fixed library of predefined anchor or backbone families. The system prompts provided to the agents contain no task-specific chemical priors, and the Blackboard is initialized empty and expands throughout the campaign as the agents introduce new ligands and anchor definitions (Figure S2).

All proposed molecules pass through a programmatic admission gate before entering a fixed atomistic evaluation pipeline. Accepted ligands are placed at a surface Br vacancy (i.e., above an undercoordinated $Pb^{2+}$ atom) on a periodic $CsPbBr_3$(001) slab, achieving 25% surface coverage. The effects of varying ligand coverage and lateral ligand-ligand interactions were not systematically examined. The ligand-bound systems were then relaxed with a pretrained MLIP with explicit dispersion correction, which acts as a surrogate model for DFT calculations, and were scored by their binding energies. Multiple ligand placements are evaluated to account for variability in passivation geometry, and the resulting binding statistics together with the coordination geometry are returned to the Blackboard. The accuracy of the MLIP surrogate is illustrated in Figure S1. Further details on atomistic model construction, dispersion-corrected MLIP, and binding energy calculation are provided in Methods.

The Decider may also choose to derive a structure–binding design principle from the accumulated observations through a two-stage procedure. First, during *Formulate*, the Decider proposes a design principle in the form of a SMARTS-based structural predicate that operationalizes the proposed structural feature into a machine-readable molecular pattern (for example, the presence of a C=C double bond). This allows the existing pool of molecules to be partitioned into those that satisfy the predicate and those that do not, and the binding-energy effect associated with the predicate-defined features can be evaluated statistically. Principles that pass this retrospective screening are designated as candidate principles but are not yet considered prospectively supported. In a subsequent *Test,* the LLM designs matched molecular pairs[33] intended to differ only in that structural feature. A principle is promoted to a supported principle only when the predicted binding-energy effect is reproduced prospectively in these newly designed molecules and satisfies the predefined statistical criteria. Every proposed principle is additionally subjected to a programmatic confound audit that evaluates whether the observed effect may be associated with other variables such as anchor identity or molecular size. These diagnostics are reported to the Decider but do not alter the formal verdict of the principle.

Throughout the campaign, the Summarizer distills the accumulated campaign outcomes into a compact playbook of successful, unsuccessful, and unresolved molecular modifications. This playbook, together with the current ligand pool and design principle ledger, is written back to the Blackboard and made available to the Decider in subsequent iterations. Through this evolving shared context, PerovAgent progressively adapts both its molecular search strategy and its chemical hypotheses *via* in-context learning. The resulting closed loop therefore couples molecular optimization with explicit hypothesis formulation and validation, allowing the search trajectory itself to become a source of testable structure–binding design principles.

## Search behavior of PerovAgent

We examined how PerovAgent navigates the ligand design space over the course of a discovery campaign. Ten independent campaigns were performed under the same search policy, with five allowing ligand-bound halogens and five excluding them. This split was introduced because preliminary searches revealed a systematic bias towards halogen-rich regions of chemical space, as discussed further below. In total, the ten campaigns comprised 198 ReAct iterations and generated 1,426 distinct ligands. Rather than following a predefined schedule, PerovAgent selected its actions adaptively throughout each campaign. Across all runs, 132 iterations (66.7%) were devoted to *Experiment*, comprising 26 *Explore* and 106 *Exploit*; 63 (31.8%) to *Derive Principle*, comprising 35 *Formulate* and 28 *Test*; and 3 (1.5%) to Stop (Figure 2a). A representative trajectory further illustrates that *Explore*, *Exploit*, *Formulate*, and *Test* remain interleaved throughout the search, with exploration often revisited even after several rounds of exploitation (Figure 2c). This behavior illustrates that PerovAgent can return to broader sampling or hypothesis testing when necessary as the campaign context evolves.

Ligand binding energy improved progressively over the course of the campaigns. Averaged across the ten runs, the mean binding energy of the accumulated pool became gradually more negative with iteration. The mean campaign-best binding energy improved from −3.59 eV after the initial iteration to −4.64 eV near the end of the search (Figure 2b). In every campaign, the final best-performing ligand emerged through a multi-step modification lineage originating from a *de novo Explore* or *Test* proposal (Figure 2c). These lineages contained two to ten successive child generations (mean, 6.7) and achieved a median binding-energy improvement of 0.78 eV from the initial candidate to the final winner. These trajectories indicate that much of the optimization arose from iterative refinement of promising ligands identified earlier, rather than from single-shot generation of high-performing molecules.

Beyond optimizing individual ligands, PerovAgent also incorporated the derived structure–binding principles into subsequent ligand proposals. To assess whether the derived principles influenced later search behavior, we compared the fraction of the molecular pool satisfying each structural predicate before and after its derivation (Figure 2d). Predicates corresponding to supported principles increased in prevalence by an average of 51.3 percentage points (pp), whereas weak principles increased by a modest 21.8 pp and refuted principles decreased by 3.4 pp. Thus, supported principles were preferentially reinforced in subsequent proposals, while refuted principles showed no comparable enrichment. This shift in pool composition is consistent with the shared campaign context influencing subsequent molecular generation through in-context learning.

Figure 2e shows the strongest-binding ligands identified in each campaign. All ten contained either phosphonate or phosphate anchors, consistent with previous experimental and computational reports of their strong binding to halide perovskite nanocrystal surfaces,[11,34–37] while retaining diverse molecular backbones. More broadly, PerovAgent sampled 1,426 distinct ligands spanning 105 Murcko scaffolds (Figure S3), demonstrating substantial chemical diversity across campaigns. The halogen-allowed campaigns favored highly halogenated structures, whose ligand-bound halogens frequently formed additional interactions with surface $Cs^+$ ions (Figure S4a). Consistent with this observation, ligand binding energy correlated with the number of ligand-bound halogen atoms (Figure S5). However, no halogenated ligands appeared on the Pareto front when binding energy and synthetic accessibility (SA) were considered jointly (Figure S6). The strongest-binding ligand identified across all ten campaigns adopted a surface-proximal conformation in which a ring nitrogen and two oxygen atoms in the substituted side chain formed additional contacts with surface $Cs^+$ ions (Figure S4b), while retaining a favorable synthetic accessibility score. Overall, PerovAgent demonstrated adaptive search behavior, gradually optimizing the binding-energy objective while navigating a broad ligand design space through reasoning-guided exploration, refinement, and principle-driven feedback.

## From local reasoning to design principles

Optimization performance alone does not establish whether the chemical reasoning generated during the search is scientifically reliable. Because each ligand proposal is accompanied by a mechanistic hypothesis and a predicted direction of binding-energy change before the atomistic evaluation, the campaign provides a direct record of both successful and unsuccessful reasoning. We therefore evaluated this reasoning at two levels: whether local hypotheses correctly predict the direction of individual molecular modifications, and whether higher-level design principles derived from recurring observations remain supported under prospective testing and confound auditing.

LLM-generated hypotheses contained useful directional information, although raw accuracy was strongly affected by the class imbalance of the optimization landscape (Figure 3a). Across all scoreable molecular modifications, PerovAgent achieved a raw direction-prediction accuracy of 0.80. However, 88% of evaluated modifi-cations weakened binding, meaning that a naïve predictor that always predicts weaker binding would achieve an even higher raw accuracy of 0.88. When we used balanced accuracy, defined as the mean of sensitivity and specificity (see Methods), to account for this imbalance, PerovAgent achieved a balanced accuracy of 0.68 against the always-weaker baseline of 0.50. Thus, the agent's local hypotheses contained meaningful predictive signal beyond the dominant tendency for structural modifications to weaken binding, although they were not uniformly correct.

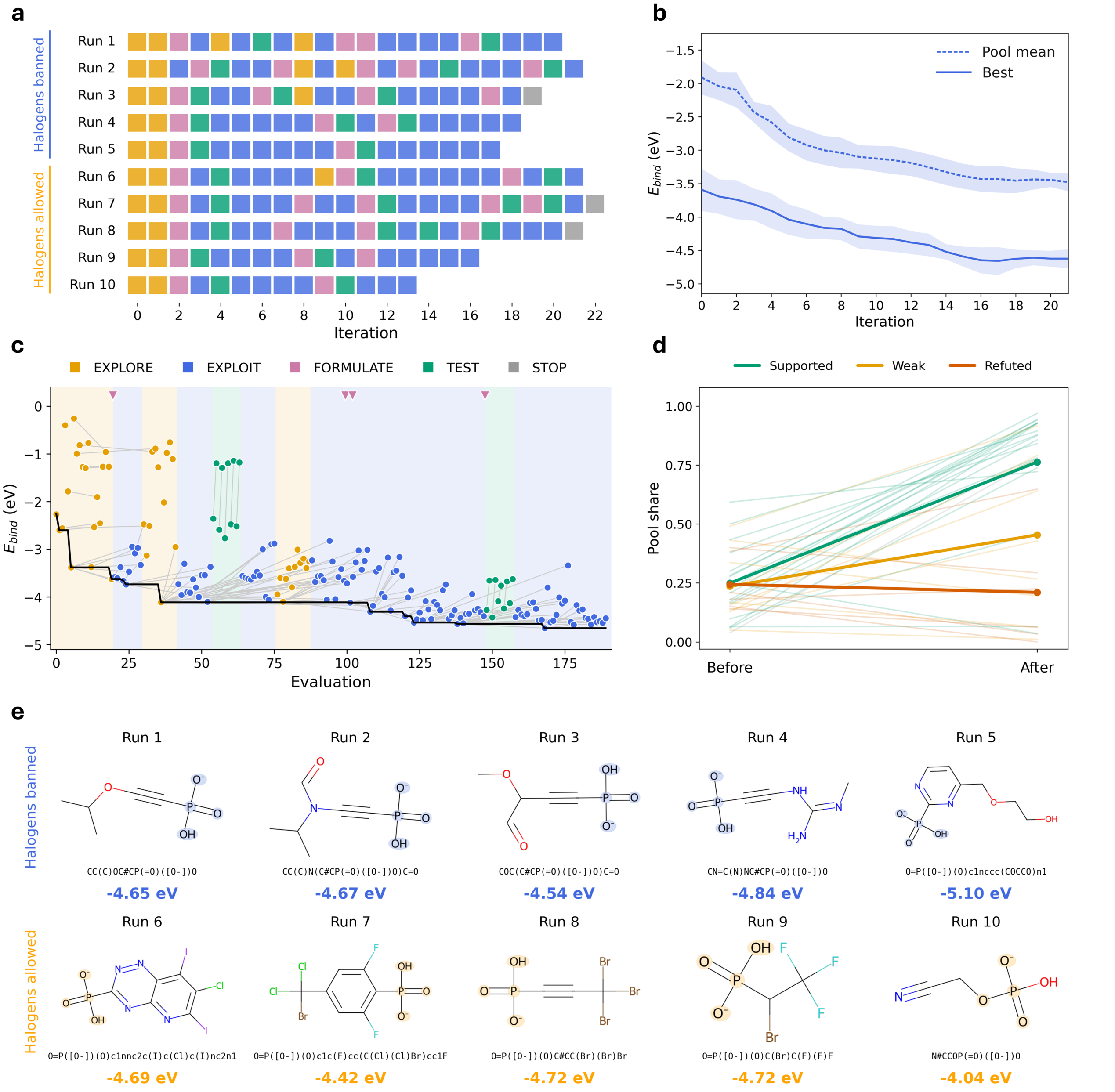

**Figure 2: Adaptive search behavior of PerovAgent across ten independent campaigns. a)** Action sequences for the ten campaigns, with Runs 1-5 conducted under a halogen-banned policy and Runs 6-10 under a halogen-allowed policy. Colors denote *Explore*, *Exploit*, *Formulate*, *Test*, and *Stop* actions. **b)** Evolution of the pool-mean and campaign-best binding energies over iterations, aggregated across the ten campaigns. The shaded regions represent ±1 standard deviation. **c)** Representative search trajectory from Run 1. Points show evaluated ligands colored by the action that generated them; gray lines connect parent-child molecular modifications, and the black line tracks the best binding energy attained up to each evaluation. Triangles indicate *Formulate* actions. **d)** Change in the fraction of the pool satisfying each proposed structural predicate before and after principle derivation, grouped by the final verdict (supported, weak, or refuted). Thin lines represent individual principles, and thick lines show the group average. **e)** Structures and binding energies of the best ligand identified in each campaign, shown separately for the halogen-banned and halogen-allowed runs.

Individual examples further illustrate both the specificity and the limitations of this molecule-specific chemical reasoning (Figure 3b). Opening the cyclohexane ring of cyclohexanesulfonic acid while preserving the sulfonic acid head group was predicted to strengthen binding by relieving conformational constraints at the surface. The resulting molecule indeed bound more strongly, from −3.13 to −3.46 eV. The relaxed structure exhibited shorter Pb-O and Cs-O contacts, decreasing from 2.99 to 2.95 Å and from 3.27 to 3.18 Å, respectively, consistent with reduced steric obstruction. In contrast, replacing a methyl hydrogen of methylphosphonic acid with an amino group was predicted to introduce an additional surface interaction and strengthen binding, but instead weakened binding from −3.89 to −3.20 eV. Structural relaxation showed that the added group disrupted the Pb−O contact of the parent ligand, replacing the short monodentate contact at 2.43 Å with two longer ones at 2.54 and 2.93 Å. These examples demonstrate that the agent generates chemically specific and falsifiable hypotheses, while also showing that chemically plausible rationales are not necessarily correct. Prediction statistics by experiment mode further illustrate these limitations (Figure S7). More importantly, predictive success at the level of individual modification does not establish that the underlying chemical rationale transfers across molecules.

The explicit *Derive Principle* layer of PerovAgent addresses this transferability problem by converting recurring observations into structural predicates that are subjected to retrospective screening, prospective testing, and confound auditing. Across ten campaigns, PerovAgent formulated 35 principles, 28 of which were prospectively tested (Figure 3c). Nineteen met the prospective support criterion at $\alpha = 0.10$, and seven additionally passed the predefined confound audit for anchor identity and molecular size. A *post hoc* sensitivity analysis using a more stringent significance threshold of $\alpha = 0.05$ retained six of these seven audit-passing principles (Figure S8). Collapsing chemically equivalent predicates further reduced these to four distinct design claims. This progressive attrition shows that candidate structure–binding relationships are readily generated from campaign history, whereas substantially fewer remain supported when challenged on newly designed molecules and examined for potential confounders. Because candidate principles are formulated from and retrospectively screened on the same campaign pool, apparent associations may reflect selection effects or chance correlations[38]. Prospective testing therefore provides a substantially stronger test of whether the predicted effect generalizes beyond the data from which the principle was derived.

Recurrence across independent campaigns provides an additional measure of whether these principles capture reproducible structure–binding rather than search-specific explanations (Figure 3d). The most frequently proposed principle was the strong binding associated with phosphonic and phosphoric acid head groups, which emerged independently in nine campaigns. Although consistent with established experimental and computational observations[11,34–37], this principle largely restates anchor identity and does not constitute a backbone-level structure–binding relationship. In contrast, the presence of an alkyne directly bonded to phosphorus passed the confound audit and was independently recovered in four campaigns, with measured strengthening effects of approximately 0.23–0.69 eV. For this principle, the agent proposed that the linear C≡C linkage reduces steric obstruction around the binding group and facilitates closer access to the surface (Figure 4a). This interpretation is consistent with prior studies showing that compact ligand backbones can promote closer surface coordination.[39]

Beyond the most recurrent alkynyl-phosphonate principle, three additional distinct design principles that passed the confound audit captured other structure–binding relationships (Figure S9–S11). Prospective matched-pair tests favored direct P–aryl over P–O–aryl connectivity (mean improvement of 0.45 eV), a ring nitrogen adjacent to the phosphonate-bearing carbon over the corresponding phenyl counterpart (0.26 eV), and ortho- over meta-hydroxyl substitution in arylphosphonates (0.40 eV). These results demonstrate that the agent-derived principles extend beyond a single chemical motif, encompassing phosphorus connectivity, heteroaromatic substitution, and local functional-group positioning.

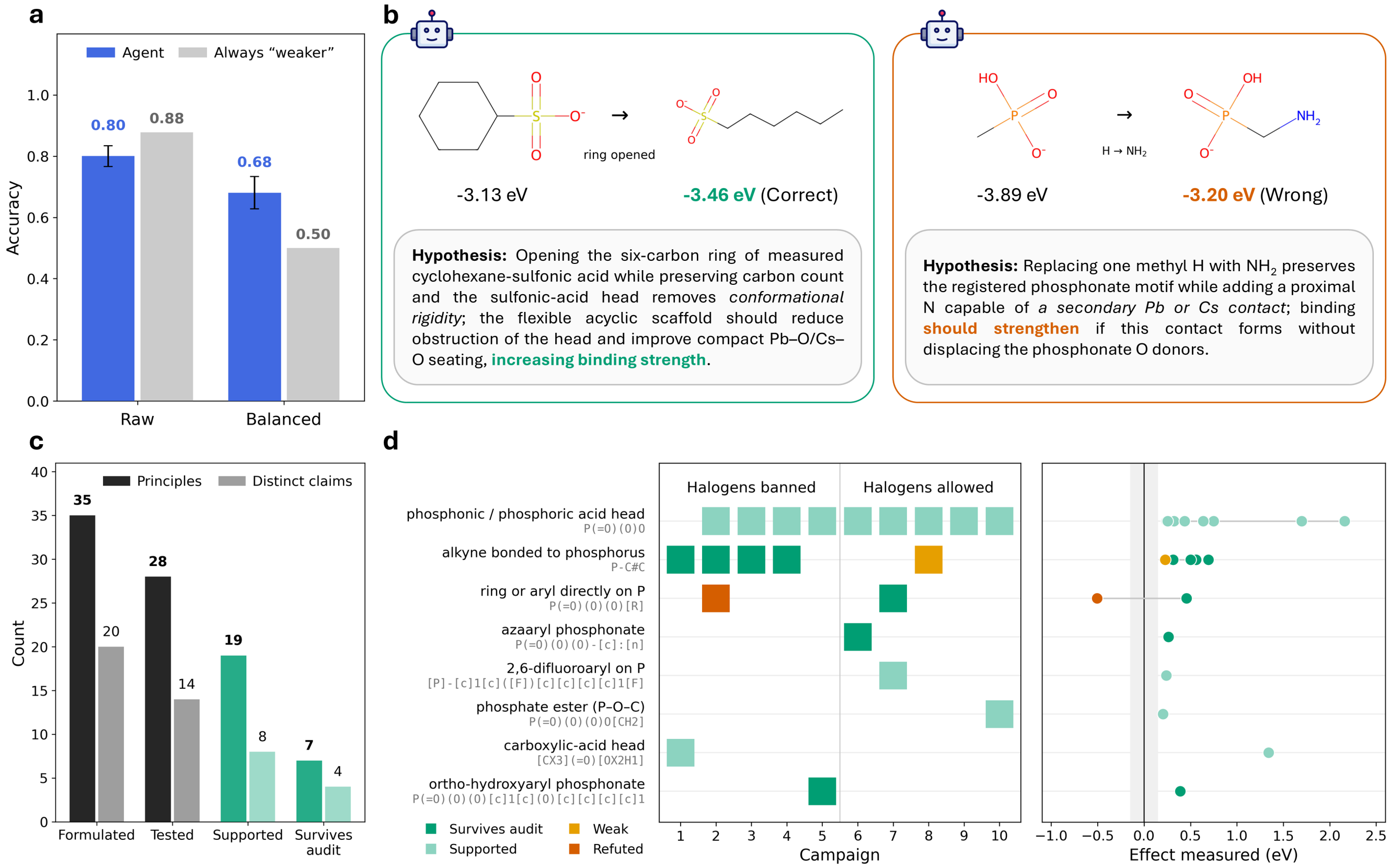

**Figure 3: Evaluation of LLM-generated scientific reasoning. a)** Raw and balanced accuracy of the agent's predicted binding-energy direction compared with an always-weaker baseline that predicts weaker binding for every modification. Error bars indicate 95% confidence intervals from a percentile bootstrap over the ten campaigns. **b)** Representative examples of correct and incorrect directional hypotheses for individual molecular modifications, showing the predicted structural rationale and the subsequently evaluated change in binding energy. **c)** Attrition of derived design principles through *Test* and confound auditing; black bars indicate the number of principles and gray bars indicate the corresponding number of chemically distinct claims. **d)** Recurrence and measured effect sizes of design principles across the ten campaigns. The left panel indicates the campaigns in which each principle was proposed and its final verdict; the right panel shows the corresponding measured binding energy effect. Dark green denotes supported principles that pass the confound audit. The gray shaded region in the right panel indicates the 0.15 eV effect-size threshold.

Together, these results distinguish two levels of scientific reasoning in PerovAgent: local hypotheses can provide useful guidance for individual molecular edits, whereas more generalizable design rules emerge only after explicit abstraction, prospective testing, and confound auditing. These supported principles can in turn inform subsequent proposals, steering the search toward regions of ligand design space enriched in prospectively supported structural features.

## Validation of the PerovAgent-derived design principle

Among these four distinct supported design principles, we selected the recurrent alkynyl-phosphonate relationship for further examination. During a *Test* round in Run 1, the LLM generated five matched molecular pairs that preserved the phosphonate anchor and heavy-atom count while differing in the degree of unsaturation of the linkage adjacent to phosphorus (Figure 4a). In all five pairs, the dispersion-corrected MLIP predicted stronger binding for the alkynyl ligand than for its alkenyl analog, with a mean improvement of 0.562 ± 0.17 eV. Importantly, the same principle was independently formulated and prospectively tested in Run 3 using alkyl rather than alkenyl controls (Figure S12). All five matched pairs again favored the alkynyl analog, with a mean binding-energy improvement of 0.694 ± 0.25 eV, and the agent independently invoked the same reduced steric obstruction rationale. This independent replication indicates that the alkynyl preference was not specific to a single campaign. This trend extended beyond these prospective test sets. Across all halogen-free phosphonate ligands sampled in the ten campaigns, increasing linker unsaturation was associated with progressively more favorable binding energies (Figure 4b). The mean improvement was 0.195 eV from alkyl to alkenyl and 0.459 eV from alkenyl to alkynyl, indicating that the alkynyl preference was reproduced across diverse molecular backbones.

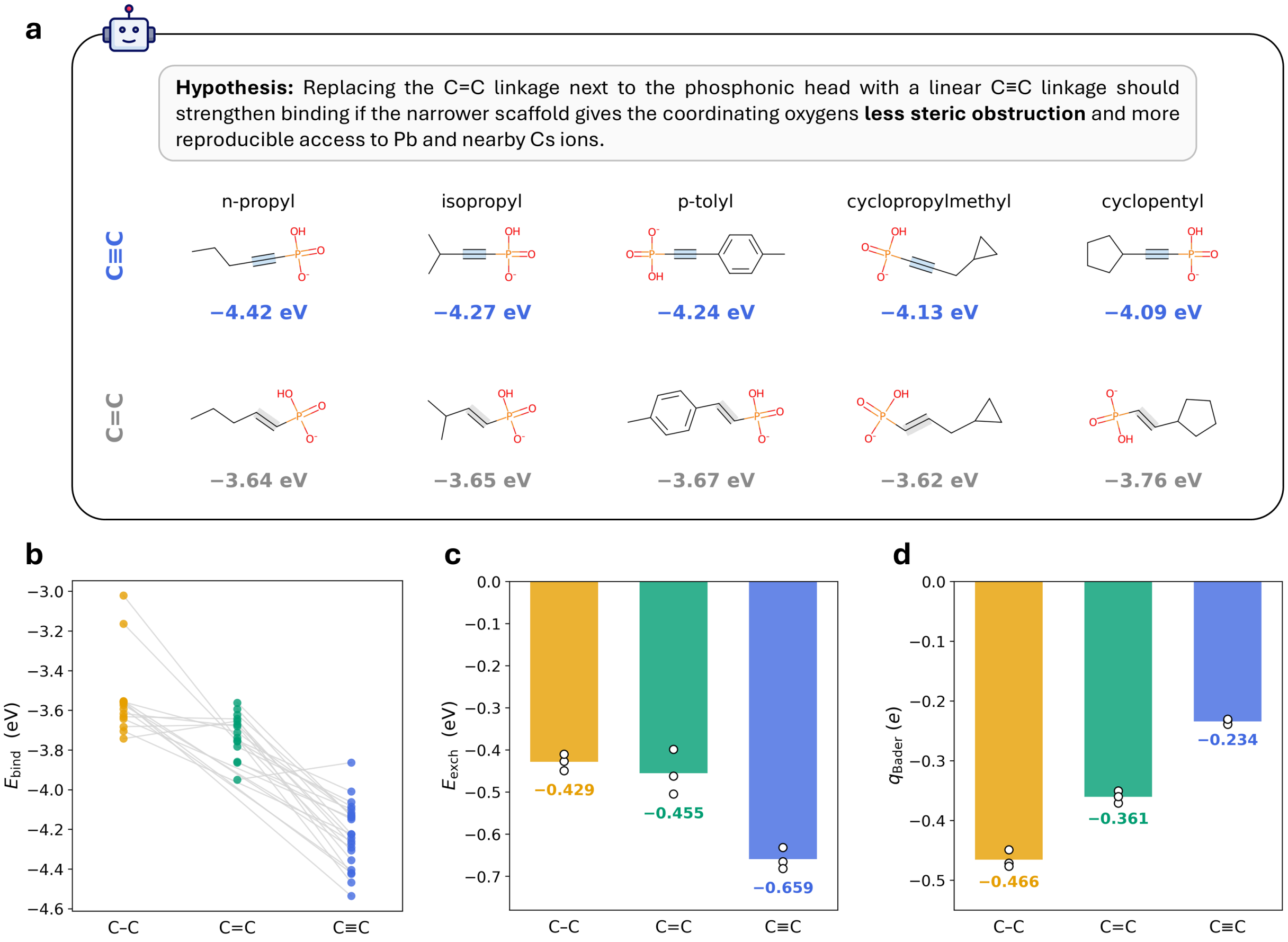

**Figure 4: DFT validation of the alkynyl-phosphonate design principle.** **a)** Five matched molecular pairs generated during *Test* in Run 1, comparing alkynyl and alkenyl linkers directly bonded to phosphorus while preserving the phosphonate anchor and heavy-atom count. **b)** MLIP-calculated binding energies of all halogen-free phosphonate ligands containing alkyl, alkenyl, or alkynyl linkages adjacent to phosphorus across the ten campaigns. Gray lines connect ligands sharing the same R substituent. **c)** DFT ligand-exchange energies, and **d)** Bader charges for the pentylphosphonate, pent-1-enylphosphonate, and pent-1-ynylphosphonate evaluated from three independent initial configurations. Bars indicate the mean ligand-exchange energy and Bader charges of the phosphonate anchors, respectively, and open circles denote individual calculations.

We next evaluated whether this design principle persisted at the DFT level for a more experimentally relevant protic-ligand binding and halide displacement on halide perovskite nanocrystal surfaces. Following previous treatments of anionic (X-type) ligand chemisorption on $CsPbBr_3$ surfaces,[40] we described binding through a ligand-exchange process involving ligand deprotonation and surface-halide displacement. Here, the ligand-exchange energy is the calculated reaction energy for replacement of a surface $Br^-$ by a singly deprotonated phosphonic acid ligand, with the displaced bromide combining with the released proton to form HBr. More negative values indicate more energetically favorable substitution. Across all X-type ligands identified in the campaigns, the ligand-exchange energy showed strong rank agreement with the campaign binding-energy metric (Spearman $\rho = 0.837$; Figure S13), indicating that the overall energetic ordering was largely robust to the choice of reference definition. We then evaluated pentylphosphonate, pent-1-enylphosphonate, and pent-1-ynylphosphonate at the PBE-D3(BJ) level. The qualitative ordering observed in the campaign was reproduced consistently with DFT (Figure 4c): all three calculations for pent-1-ynylphosphonate yielded more favorable exchange energies than those for pent-1-enylphosphonate and pentylphosphonate. The DFT-relaxed structures were also consistent with the steric explanation proposed by the agent. The van der Waals volume of the bound pent-1-ynylphosphonate was 7.9% and 4.0% smaller than those of the pentyl and pentenyl counterparts, respectively, reflecting the compact geometry of the linear *sp*-hybridized linkage (Figure S14). Consistent with reduced steric congestion around the phosphonate heads, the alkynyl substitution also allowed slightly shorter bridging Pb–O contacts of 2.722 and 2.769 Å, compared with 2.734 and 2.801 Å corresponding to those of the pent-1-enylphosphonate.

To examine whether electronic effects accompany the structural differences, we analyzed Bader charges for the same three systems. The phosphonate anchors exhibited progressively less negative Bader charges from alkyl (−0.466 e) to alkenyl (−0.361 e) and alkynyl (−0.234 e) substitution (Figure 4d), indicating a systematic redistribution of electron density associated with the linker hybridization. This trend is qualitatively consistent with the greater electron-withdrawing character of *sp*-hybridized carbon, arising from its higher *s*-character relative to $sp^2$- and $sp^3$-hybridized carbon.[41] Such electronic effects could influence the acidity of the phosphonic acid group[40] and, consequently, the energetics of ligand deprotonation involved in the ligand-exchange reaction. Together, these results suggest that electronic effects may contribute, alongside steric effects, to the observed preference for alkynyl substitution.

## Methodological implications of agentic closed-loop discovery

Our vision for agentic materials discovery is to organize computational exploration as a sequential scientific process: propose candidates using chemical knowledge, evaluate them, and use the resulting evidence to revise subsequent decisions. LLMs are well suited to this role because they can reason over molecular structures and qualitative chemical concepts without requiring predefined numerical descriptors to guide design. We first explored this philosophy in LLMatDesign[24] through direct structural modification and iterative self-reflection, and subsequently extended it in MAESTRO[25] through multi-agent collaboration for activity-stability optimization. PerovAgent advances this approach by integrating adaptive action selection, computational evaluation, explicit design-principle testing, and history summarization within a shared feedback loop (Figure 1). The progressive improvement in binding energy and the multigeneration modification lineages leading to the strong-binding ligands support the value of this iterative strategy: computational outcomes inform successive refinements, while the summarized campaign history preserves useful observations for later decisions.

The methodological contribution extends beyond assigning candidate generation and evaluation to different agents. PerovAgent makes accumulated experience available for reuse and subjects the chemical relationships inferred from that experience to explicit scrutiny. Its design-principle layer converts recurring observations into structural predicates that undergo retrospective screening, prospective testing, and confound auditing, distinguishing plausible explanations from relationships supported on newly designed molecules. The substantial attrition of proposed principles during validation (Figure 3c) demonstrates the need for these safeguards, while the preferential enrichment of supported structural features in subsequent proposals (Figure 2d) provides behavioral evidence that tested principles feed back into the search. Together, these results support our broader philosophy that agentic discovery benefits from a loop in which reasoning is continually revised against computational evidence. The advances in PerovAgent make this loop more systematic and scientifically accountable by connecting candidate optimization with the formulation, testing, and reuse of chemical knowledge.

# Conclusions

In this work, we developed PerovAgent as an autonomous framework for navigating ligand chemical space and deriving structure–binding design principles for halide perovskite surfaces. Across ten independent campaigns, the framework evaluated 1,426 distinct ligands and progressively improved the binding-energy objective while maintaining substantial diversity in molecular scaffolds and binding motifs. Beyond optimization, the campaign trajectories provided a direct record of the chemical reasoning underlying each ligand proposal. Of 35 formulated principles, 19 were prospectively supported, and seven additionally passed the predefined confound audit, underscoring the need for systematic testing to distinguish plausible hypotheses from more robust structure–binding design principles. For the recurrent alkynyl-phosphonate principle, DFT calculations reproduced the predicted energetic preference. Together, these results demonstrate an autonomous discovery framework that couples molecular optimization with scientific reasoning and systematic validation of LLM-generated design principles. More broadly, this approach provides a route toward autonomous materials discovery systems that identify promising candidates while transforming their search trajectories into falsifiable, systematically testable, and interpretable design principles.

# Methods

## Large language models

LLM agents were implemented with Pydantic AI[42] (v2.32.1) using its `OpenAIResponsesModel` interface and the `gpt-5.6-sol` model[43]. Model parameters such as reasoning effort, temperature and top-*p* were set to the default values. The three agents (Decider, Proposer, and Summarizer) were instantiated separately with distinct system prompts and output schemas to serve their respective roles throughout the campaign (See Table S3–S7). The language model's response generation was non-deterministic and not seeded; therefore, each campaign run was treated as an independent run, even when the same campaign policy is used. The agents were not provided with web access, retrieval tools, or external knowledge bases. Consequently, chemical domain knowledge available to the agents was limited to the model's pretrained knowledge and the information explicitly provided through the system prompts and evolving campaign context only. Token usage across the campaigns is shown in Figure S15.

## Explored ligand design space

The design space was restricted to monovalent anionic (X-type) and neutral (L-type) ligands[44] capable of coordinating to surface Pb atoms. Ligands were composed of H, C, N, O, S, and P, with F, Cl, Br, and I additionally permitted in the halogen-allowed campaigns. Only ligands containing 3–50 heavy atoms were considered. The admission gate screened all proposed ligands for compliance with these constraints, as well as for basic chemical validity and duplicate molecular structures.

## Atomistic model construction

PerovAgent requires an efficient and robust pipeline for constructing chemically reasonable atomistic models across the diverse molecular inputs encountered during discovery. We developed a custom Python library, PerovKit, for the programmatic construction of atomistic models of ligand-passivated halide perovskite slabs and isolated molecules. A pristine cubic $CsPbBr_3$ slab of 2×2×3 unit cells is first constructed with a lattice parameter of 5.95 Å, periodic in the *x* and *y* directions and with 20 Å of vacuum along *z*. Both surfaces are terminated with Cs and Br atoms, so the slab (68 atoms, $Cs_{16}Pb_{12}Br_{40}$) is formally charge-neutral. The SMILES[29] provided by the Proposer is kept neutral for an L-type ligand or deprotonated at the proposed anchor for an X-type ligand.

The ligand is then instantiated from the resulting SMILES and occupies one of the four surface Br sites substitutionally (25% X-site coverage) such that the proposed binding motif (one or two select atoms) coordinates the Pb atom directly below. The placement algorithm assigns the ligand a random initial orientation about the surface normal (+z) and then rotates it about that axis until the shortest distance between any ligand atom and any slab atom exceeds a fixed cutoff of 1.6 Å. The algorithm is sufficiently robust, with slab construction failing for only 30 of the 1,971 molecules that passed the admission gate (1.5%) across the ten campaigns (Figure S16).

## Machine-learned interatomic potential

`MatterSim-v1.0.0-5M`[45] was selected as a surrogate for density functional theory (DFT) calculations. MatterSim is a graph neural network based on the M3GNet architecture[46], pretrained on a large PBE-level DFT dataset comprising bulk inorganic materials, molecular crystals, organic polymers, metal-organic frameworks, two-dimensional materials, surfaces, and interfaces across a wide range of temperatures and pressures.[45] The performance of `MatterSim-v1.0.0-5M` was benchmarked against the DFT reference test set, comprising geometry-relaxation trajectories of various ligand-passivated halide perovskite slabs from our previous study.[11] It demonstrated reliable accuracy without any fine-tuning, with a systematic energy offset associated with the absent dispersion terms (Figure S17). Therefore, we additively applied Grimme-D3 dispersion corrections[47] with Becke-Johnson (BJ) damping[48] to MatterSim-predicted energy and atomic forces via NVIDIA ALCHEMI Toolkit-Ops (v0.3.1).[49] The D3 parameters for PBE functionals were employed (a1 = 0.4289, a2 = 4.4407, s8 = 0.7875, cutoff = 20.0 Å). The final performance of MatterSim with dispersion correction is shown in Figure S1. Geometry relaxations were performed with Atomic Simulation Environment[50] (ASE, v3.28.0) using the L-BFGS algorithm[51] until all atomic forces fell below 0.025 eV/Å or a maximum of 1,000 steps was reached.

## Binding energy metric and binding geometry

The binding energy of the ligand was calculated from the MLIP-relaxed geometries of three reference systems as $E_{\text{bind}} = E_{\text{full}} - \left(E_{\text{vacancy}} + E_{\text{ligand}}\right)$, where $E_{\text{full}}$ is the energy of the ligand-passivated slab, $E_{\text{vacancy}}$ is the energy of the same slab with the ligand removed (i.e., with a single Br vacancy), and $E_{\text{ligand}}$ is the energy of the isolated ligand in its bound form. A more negative $E_{\text{bind}}$ indicates stronger binding. $E_{\text{vacancy}}$ is a common reference to every molecule in a run and is therefore computed once and cached. To obtain a low-energy conformer for calculating $E_{\text{ligand}}$, 200 ligand conformers were generated using RDKit[52] with ETKDGv3[53,54] and ranked by the universal force field[55] (UFF) energy. The 32 lowest were relaxed with MLIP and the conformation with the lowest energy was chosen. Conformers whose bond connectivity changed during relaxation were discarded. For each proposed ligand, five slabs were constructed and relaxed independently with different placement seeds. $E_{\text{bind}}$ was reported as the average, with the standard deviation across placements as placement-dependent variability. Binding geometry was characterized using the lowest energy of the five placements as a contact map, showing the distances of every ligand heteroatom within 3.8 Å of a surface Cs, Pb, or Br atom. The proposed ligand was only registered if: (1) at least three of the five placements yielded a valid structure and finite energy, (2) the standard deviation of $E_{\text{bind}}$ across placements was below 1 eV, (3) $E_{\text{bind}}$ was not more negative than -6 eV, (4) the ligand connectivity graph did not change, which otherwise indicates a surface reaction, and (5) anchor donor reached a Pb atom within an element-specific cutoff. A predefined minimum effect-size threshold of 0.15 eV was applied to MLIP-calculated binding-energy differences throughout this work, with smaller effects considered unresolved. The ligand-exchange energy[40] was computed as $\Delta E_{\text{exch}} = (E_{\text{slab}\cdot\text{L}} + E_{\text{HBr}}) - (E_{\text{slab}\cdot\text{Br}} + E_{\text{HL}})$, where $E_{\text{HL}}$ is the energy of the neutral protic ligand, $E_{\text{slab}\cdot\text{Br}}$ is the energy of the pristine surface, $E_{\text{slab}\cdot\text{L}}$ is the energy of the ligand-passivated surface, and $E_{\text{HBr}}$ is the energy of hydrogen bromide.

## Density functional theory

To validate selected campaign results at the level of density functional theory, the Vienna Ab initio Simulation Package[56] (VASP 6.4.3) was used with a projector-augmented-wave (PAW) method[57], the PBE functional[58], and Grimme's D3 dispersion correction[47] with Becke-Johnson damping[48]. The plane-wave cutoff was 400 eV (`PREC = Accurate`) with Gaussian smearing of 0.05 eV. Brillouin-zone sampling used a 3×3×1 Monkhorst-Pack *k*-point mesh[59] for the slabs and the Γ point for the isolated molecules, which were placed in an identical 20 Å cubic box. DFT slab models were constructed using the same protocol used in the production campaign runs. For geometry optimization, an electronic convergence criterion of $1\times10^{-6}$ eV was applied. The bottom three atomic layers (one $CsPbBr_3$ unit) were fixed, and all other atoms were relaxed until the forces fell below 0.025 eV/Å; slab relaxations used real-space projection and molecular calculations reciprocal-space projection. Spin-orbit coupling was not included, as its effect on geometry relaxation is negligible. Bader charges were obtained from separate single-point calculations on the converged geometries, with the real-space projection disabled and the all-electron density written. FFT grids were set explicitly and identically within each class of system (192×192×768 for slabs, 320×320×320 for molecules) to minimize differences arising from grid discretization. The charge density was partitioned with the grid-based algorithm of the Henkelman group.[60,61] Atomic charges were computed as $q = Z_\mathrm{i} - N_\mathrm{i}$, where $Z_\mathrm{i}$ is the number of valence electrons and $N_\mathrm{i}$ is the integrated charge inside the Bader volume; negative values denote electron-rich atoms. The Bader charge of the phosphonate anchor was calculated by summing the atomic charges of the bound $–PO_3H$ group (P, three O atoms, and the hydroxyl H atom).

## Statistical analysis

Over the course of a campaign, the evaluated pool comes to contain many structurally similar molecules differing only by small local modifications. As such, treating each molecule as an independent observation could overstate statistical significance. To account for this, during *Formulate*, molecules are first partitioned into two groups according to whether they satisfy or violate the structural predicate specified by a proposed design principle. Structurally similar molecules within each group are then clustered using Butina clustering[62] of Morgan fingerprints[63] (radius 2, 2048 bits) with Tanimoto similarity cutoff of 0.6. The median binding energy of each clustered unit is then regarded as an independent observation. The effect size is defined as the difference between the medians of these unit-level values for the two groups, with its sign oriented according to the predicted direction. A proposed principle is retrospectively evaluated against the existing pool using a one-sided Mann–Whitney $U$ test on the unit-level observations, requiring at least five molecules and two units per group. When both groups contain at least four units, the units are additionally divided once at random into two halves, each of which must reproduce the predicted direction of the effect. A principle is refuted if the observed effect opposes the predicted direction. It is classified as a candidate if the effect exceeds 0.15 eV in the predicted direction, yields $p < 0.10$, and, when applicable, reproduces the predicted direction in both halves; otherwise, it is classified as weak.

During *Test*, a candidate principle is prospectively evaluated using newly designed matched molecular pairs[33] intended to differ only in the predicate-defined feature. At least five molecules satisfying and five violating the predicate are required. Resolved pairs are evaluated using a one-sided Wilcoxon signed-rank test on their paired binding-energy differences. A principle is supported if the mean signed paired difference exceeds 0.15 eV in the predicted direction and $p < 0.10$. It is refuted if the observed effect opposes the prediction and classified as weak otherwise.

Separately, the agent's directional prediction performance was evaluated using raw accuracy and balanced accuracy[64], calculated as follows:

$$\text{Raw accuracy} = \frac{\mathrm{TP} + \mathrm{TN}}{\mathrm{TP} + \mathrm{TN} + \mathrm{FP} + \mathrm{FN}}$$

$$\text{Balanced accuracy} = \frac{\text{Sensitivity} + \text{Specificity}}{2}$$

$$\text{Sensitivity} = \frac{\text{TP}}{\text{TP} + \text{FN}}$$

$$\text{Specificity} = \frac{\text{TN}}{\text{TN} + \text{FP}}$$

where TP, TN, FP, and FN represent True Positives, True Negatives, False Positives, and False Negatives. Confidence intervals for raw and balanced accuracy were estimated by percentile bootstrap resampling at the campaign level using 4,000 bootstrap replicates.

### Confound audit

Whenever a design principle received a verdict, it was additionally audited against all molecules evaluated up to that point in the campaign. The predicate divided these molecules into satisfying and violating groups, and the effect was defined as the difference between the two groups' median binding energies. The audit assessed four predefined criteria. (1) Partition balance: The partition was considered degenerate if fewer than 10% or more than 90% of molecules satisfied the predicate, or if either group contained fewer than three molecules. A degenerate partition failed the audit without further evaluation. (2) Anchor confounding: The effect was recomputed within each anchor family containing at least three molecules per group. Qualifying family-level effects were averaged, weighted by family size, and the criterion was satisfied if at least one family qualified and the weighted effect exceeded the 0.15 eV threshold. (3) Separability: At least one anchor family was required to contain molecules from both groups. (4) Size confounding: The same stratified analysis as in (2) was applied after dividing molecules at the campaign-median heavy-atom count. A principle passed the audit only if all four criteria were satisfied. Audit diagnostics were reported to the agent but did not alter the principle verdict.

### Computational resources and runtime

All discovery campaigns were conducted on the Perlmutter supercomputer at NERSC using NVIDIA A100 (40 GB) GPUs. Each campaign was assigned one GPU with 16 parallel evaluation workers, taking an average wall-clock time of 4.4 hours. With concurrent evaluation workers, atomistic evaluation required an effective average of 1.07 minutes per ligand, while the complete agentic loop, including LLM API calls, averaged 1.34 minutes per ligand. Subsequent DFT validation was performed separately using the GPU version of VASP 6.4.3, typically with eight A100 GPUs per calculation.

## Acknowledgements

This material is based upon work supported by the U.S. National Science Foundation under Cooperative Agreement No. DMR-2019444 for the Science and Technology Center for the Integration of Modern Optoelectronic Materials on Demand (IMOD). This research used resources of the National Energy Research Scientific Computing Center, a DOE Office of Science User Facility supported by the Office of Science of the U.S. Department of Energy under Contract No. DE-AC02-05CH11231 using NERSC award BES-ERCAP0036338.

## Author Contributions

S.C., V.F., and G.H. conceived the study. S.C. developed PerovAgent and ran the ligand-design campaigns. S.C. and Y.J.J. analyzed the search results and evaluated the proposed design principles. S.C. performed the DFT calculations to test the predicted binding preference of alkynyl phosphonates. D.S.G., V.F., and G.H. supervised the research. S.C. drafted the manuscript. All authors contributed to the discussion of the results and reviewed the manuscript.

## Data Availability

The source codes for PerovAgent are publicly available at `https://github.com/Hu-group-at-GT/PerovAgent`. PerovKit and example snippets for slab model construction are available at `https://github.com/Hu-group-at-GT/PerovKit`. The full output of the ten campaigns shown in this work and the validation DFT data set are available at Zenodo.

Online Supporting Information for

# PerovAgent: Autonomous Ligand Design for Halide Perovskite Nanocrystals

Seungjun Cha,[1,2] Yoo Jae Jeong,[3,4] Amanda Conde Del Moral,[1] Juan-Pablo Correa-Baena,[1,5] Seth R. Marder,[3,4,6,7] David S. Ginger,[8,*] Victor Fung,[2,*] Guoxiang Hu[1,5,*]

[1] School of Materials Science and Engineering, Georgia Institute of Technology, Atlanta, GA 30332, United States
[2] School of Computational Science and Engineering, Georgia Institute of Technology, Atlanta, GA 30332, United States
[3] Materials Science and Engineering Program, University of Colorado Boulder, Boulder, CO 80309, United States
[4] RASEI; The CU-Boulder/NLR Energy Institute, University of Colorado Boulder, Boulder, CO 80309, United States
[5] School of Chemistry and Biochemistry, Georgia Institute of Technology, Atlanta, GA 30332, United States
[6] Department of Chemistry, University of Colorado Boulder, Boulder, CO 80309, United States
[7] Department of Chemical and Biological Engineering, University of Colorado Boulder, Boulder, CO 80309, United States
[8] Department of Chemistry, University of Washington, Seattle, WA 98195, United States

* *Corresponding authors*
Email: dginger@uw.edu, victorfung@gatech.edu, emma.hu@mse.gatech.edu

# Supplementary Figures

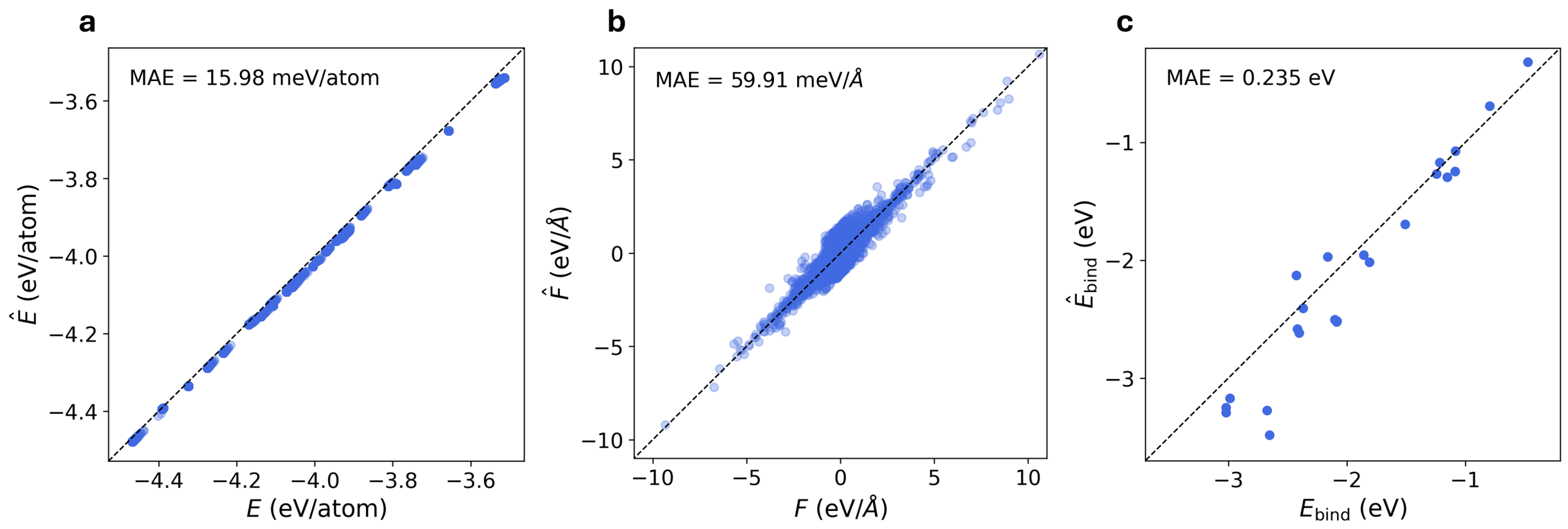

**Figure S1: Validation of the dispersion-corrected MLIP against DFT.** Parity plots of MLIP-predicted and DFT-calculated **(a)** energies, **(b)** atomic forces, and **(c)** ligand binding energies, based on the DFT reference test set of various ligand-passivated $CsPbBr_3$ slabs.

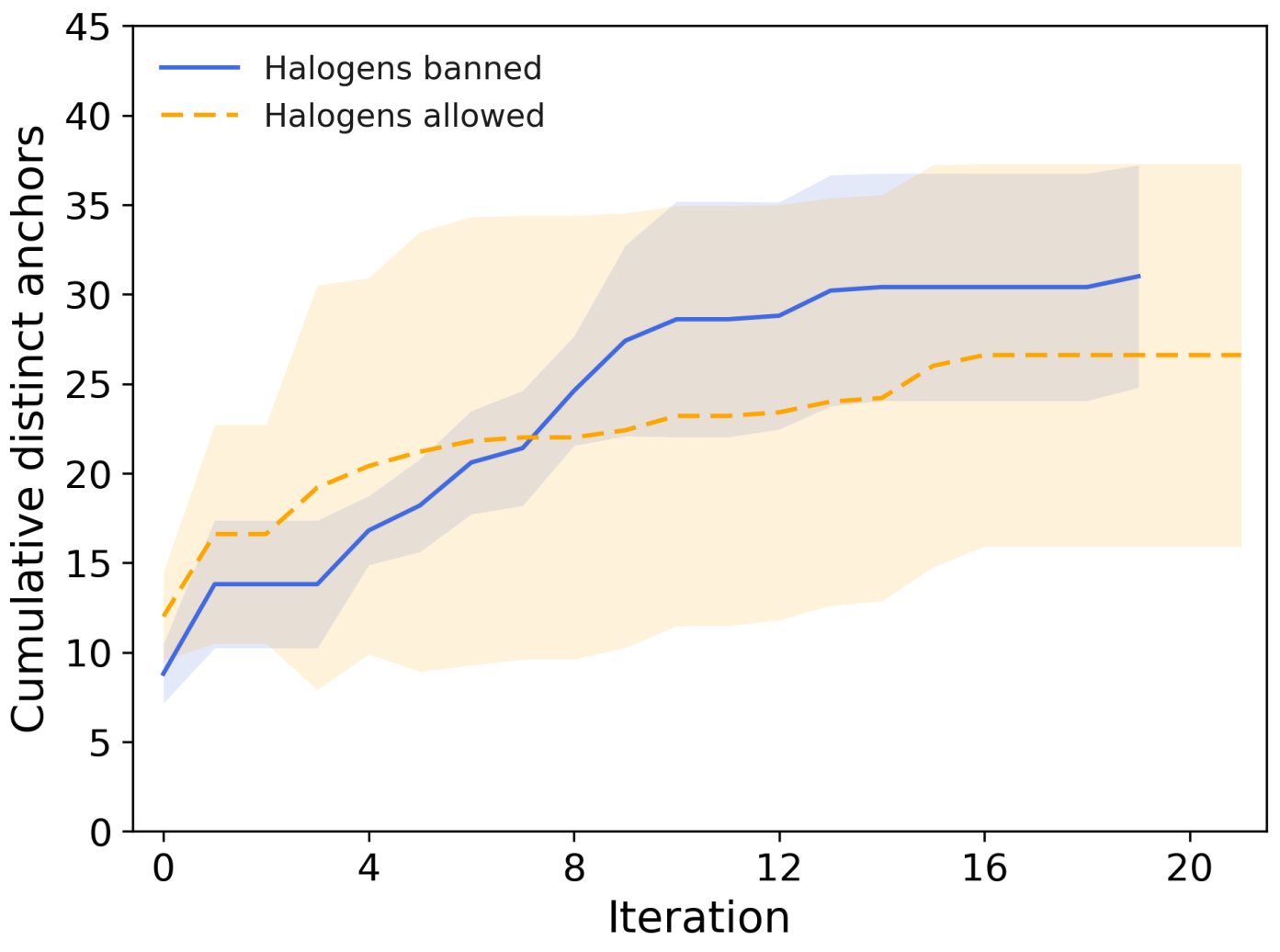

**Figure S2:** The cumulative number of distinct anchors proposed by the LLM over iteration, recorded across five halogen-banned and five halogen-allowed campaigns separately. The anchor registry is initiated with no prefilled entries and only grows as the LLM suggests a new anchor definition. 17–44 anchor names were registered per campaign, with 206 distinct names across all ten.

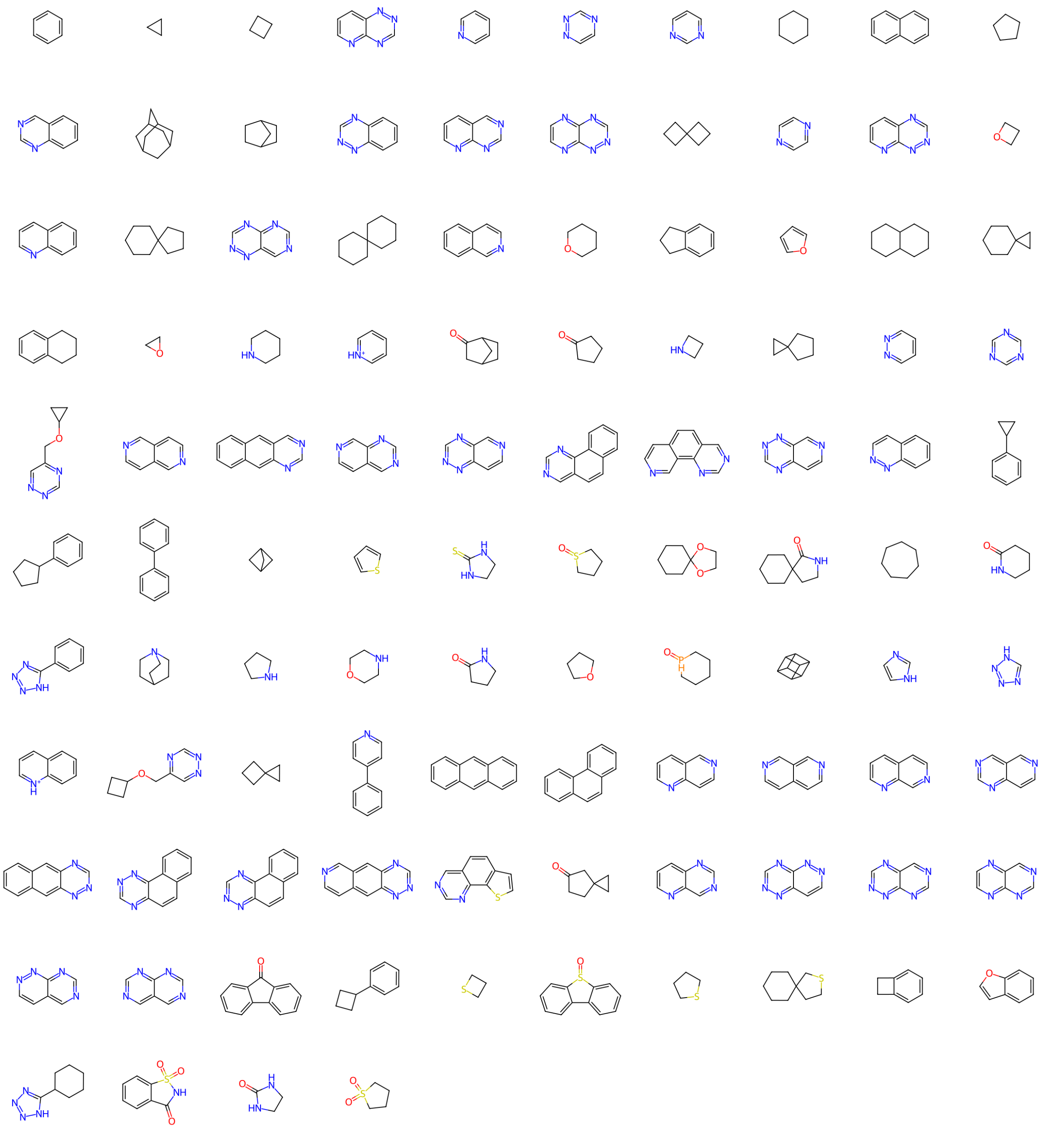

**Figure S3:** Murcko scaffolds of the 1,426 distinct ligands evaluated across the ten runs. 688 ligands (48%) are acyclic and share the empty scaffold, which is not drawn. The other 104 scaffolds are shown in row-major order, starting with the scaffold carried by most ligands.

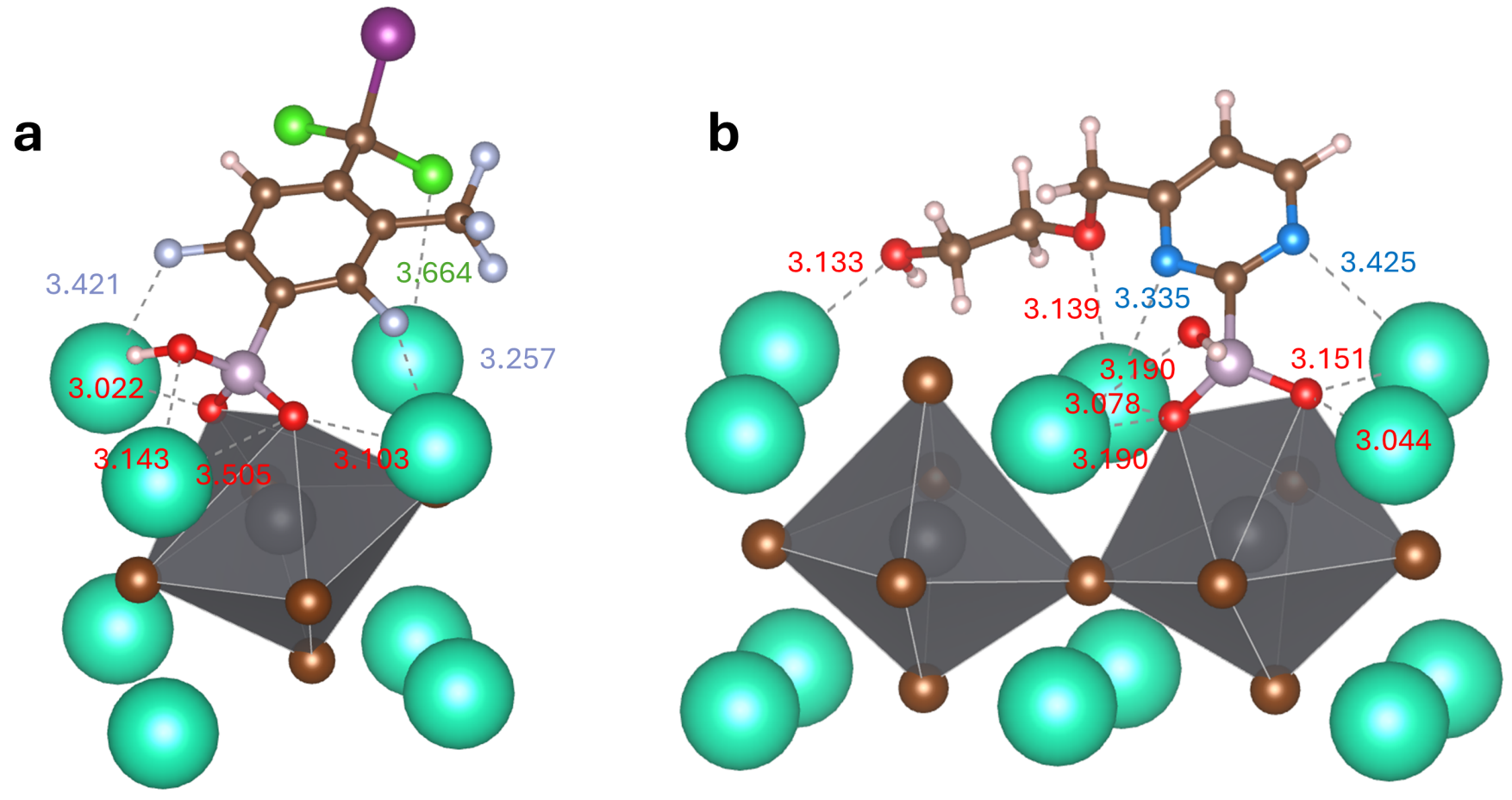

**Figure S4:** Relaxed binding configurations of **(a)** a representative halogenated ligand and **(b)** the strongest-binding ligand identified across all ten campaigns. Ligand-bound halogens form additional contacts with surface $Cs^+$ ions, providing secondary interactions that can contribute to stronger ligand binding. Similar behavior is observed for other heteroatoms (such as O and N) present in flexible side chains, which can adopt surface-proximal conformations that facilitate additional contacts with surface $Cs^+$ ions. Dashed lines indicate Cs–heteroatom contacts and numbers indicate the corresponding interatomic distances. Cs, cyan; Pb, inside gray octahedra; Br, brown; C, dark brown; O, red; P, lilac; H, light pink; N, blue; F, pale blue; Cl, green; I, purple.

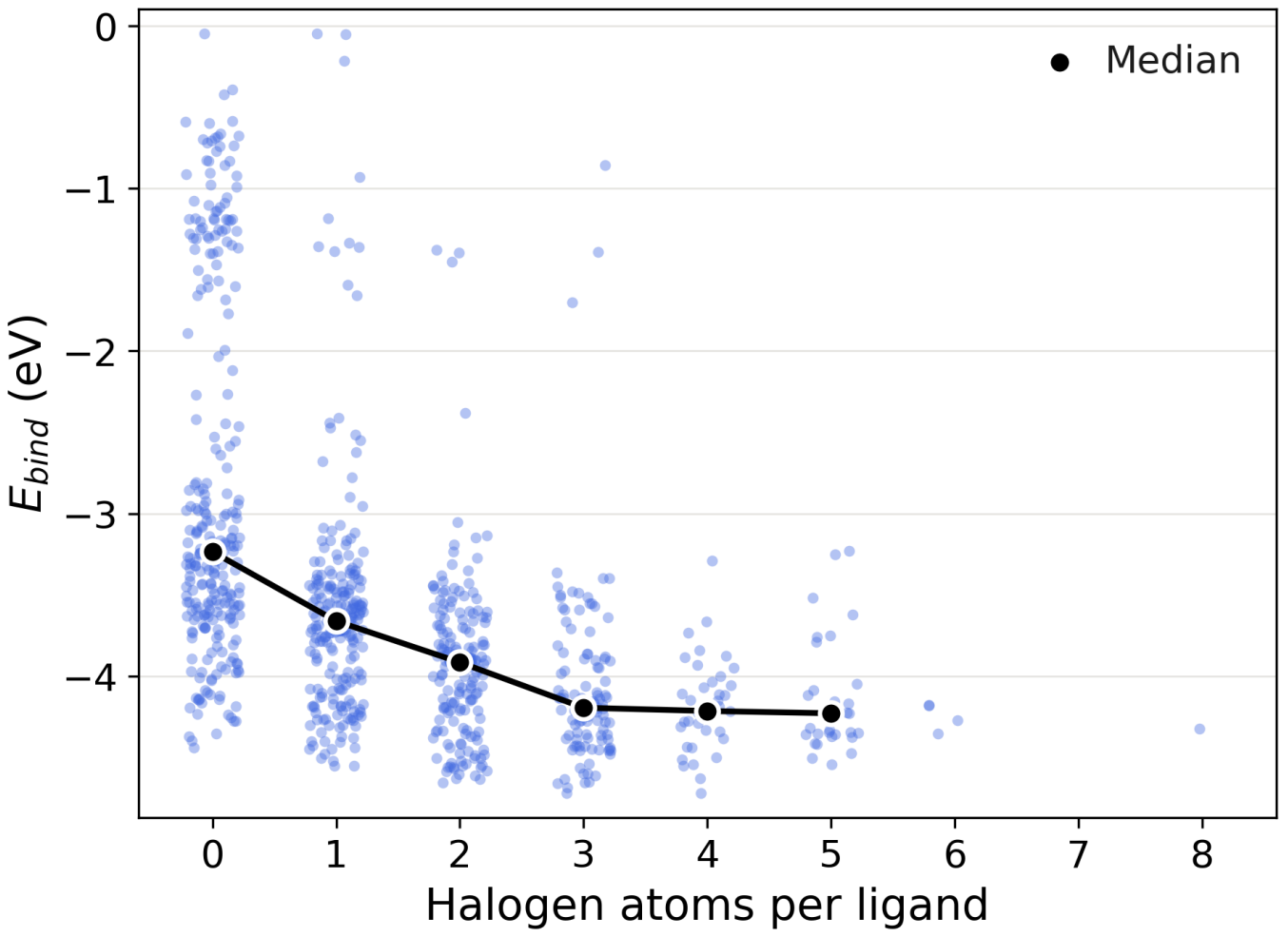

**Figure S5:** Relationship between number of ligand-bound halogens and binding energy across the five halogen-allowed campaigns. Each point represents an individual ligand, grouped by the number of ligand-bound halogen atoms. Black circles and connecting lines indicate the median binding energy within each group. Numbers in parentheses below the $x$-axis indicate the number of ligands in each group.

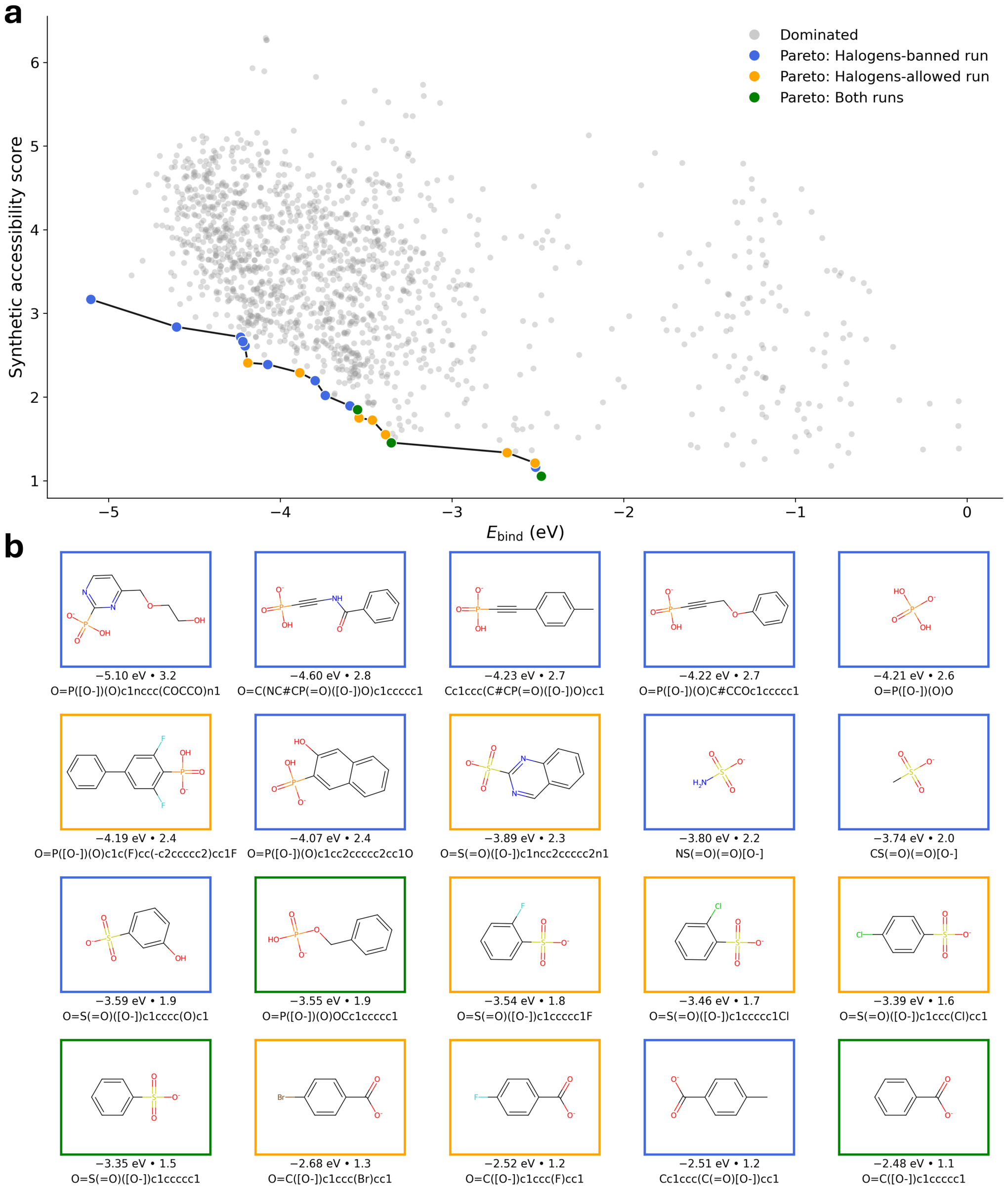

**Figure S6:** Trade-off between binding energy and synthetic accessibility (SA) score. **(a)** Binding energy against SA score for all 1,426 distinct ligands across the ten runs. A ligand evaluated more than once is plotted at its mean $E_{bind}$. SA scores are computed with the RDKit implementation of Ertl and Schuffenhauer (1 = easy to synthesize, 10 = hard) on the neutral ligand. Both quantities are minimized such that the ideal region is the lower left. Gray points represent dominated ligands, for which another ligand binds at least as strongly and is at least as easy to synthesize while being strictly better in at least one of the two objectives. The 20 non-dominated ligands that form the Pareto front are joined with a black solid line in order of $E_{bind}$, colored by the campaign policy that generated it. **(b)** The binding energy, SA score, and SMILES of the 20 Pareto-front ligands. The color of the frame represents the campaign policy, consistent with the legend in the top figure.

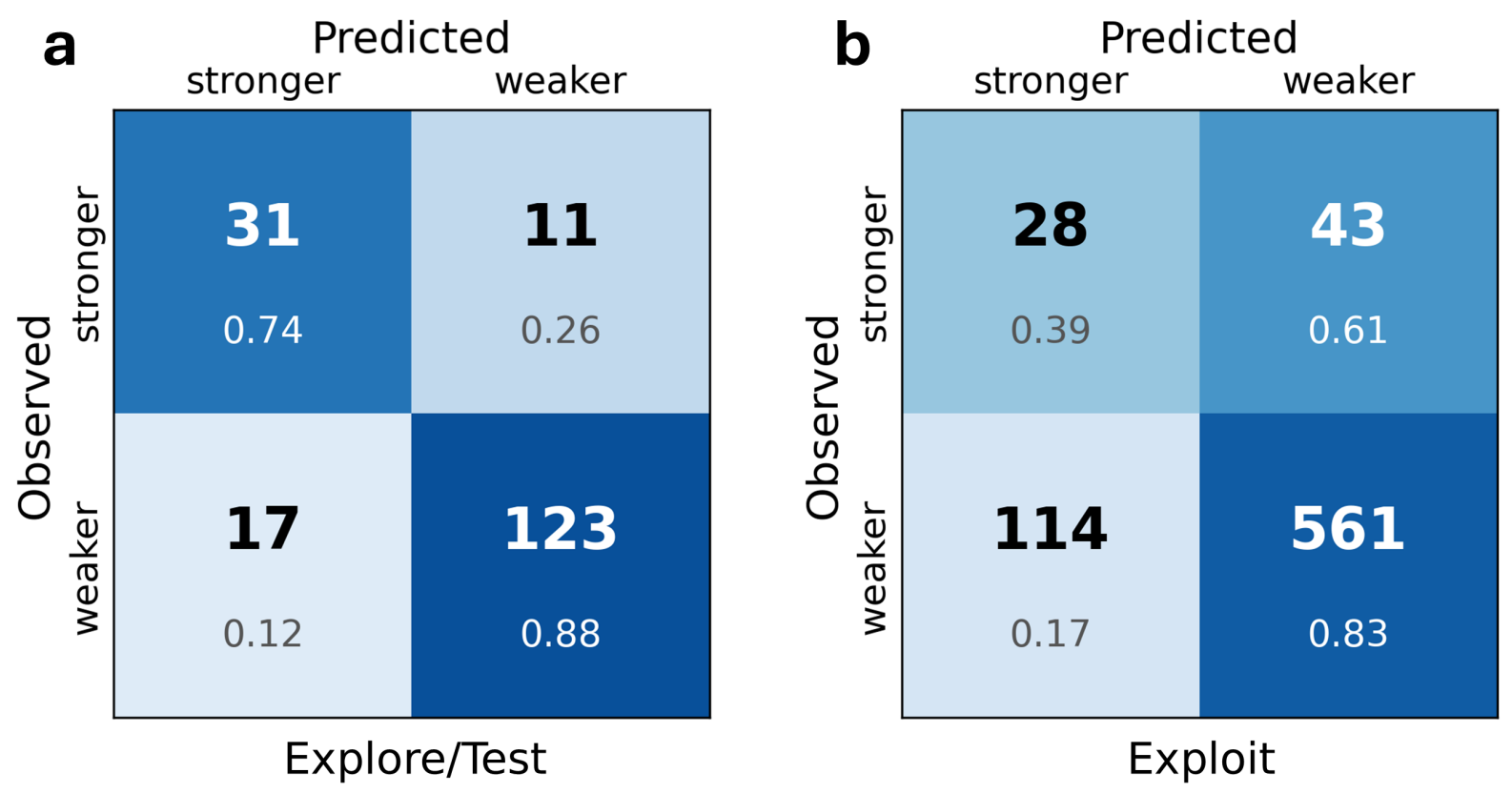

**Figure S7:** Confusion matrices for directional predictions during **(a)** *Explore/Test* and **(b)** *Exploit* rounds. Cells report the number of predictions, with row-normalized fractions shown below.

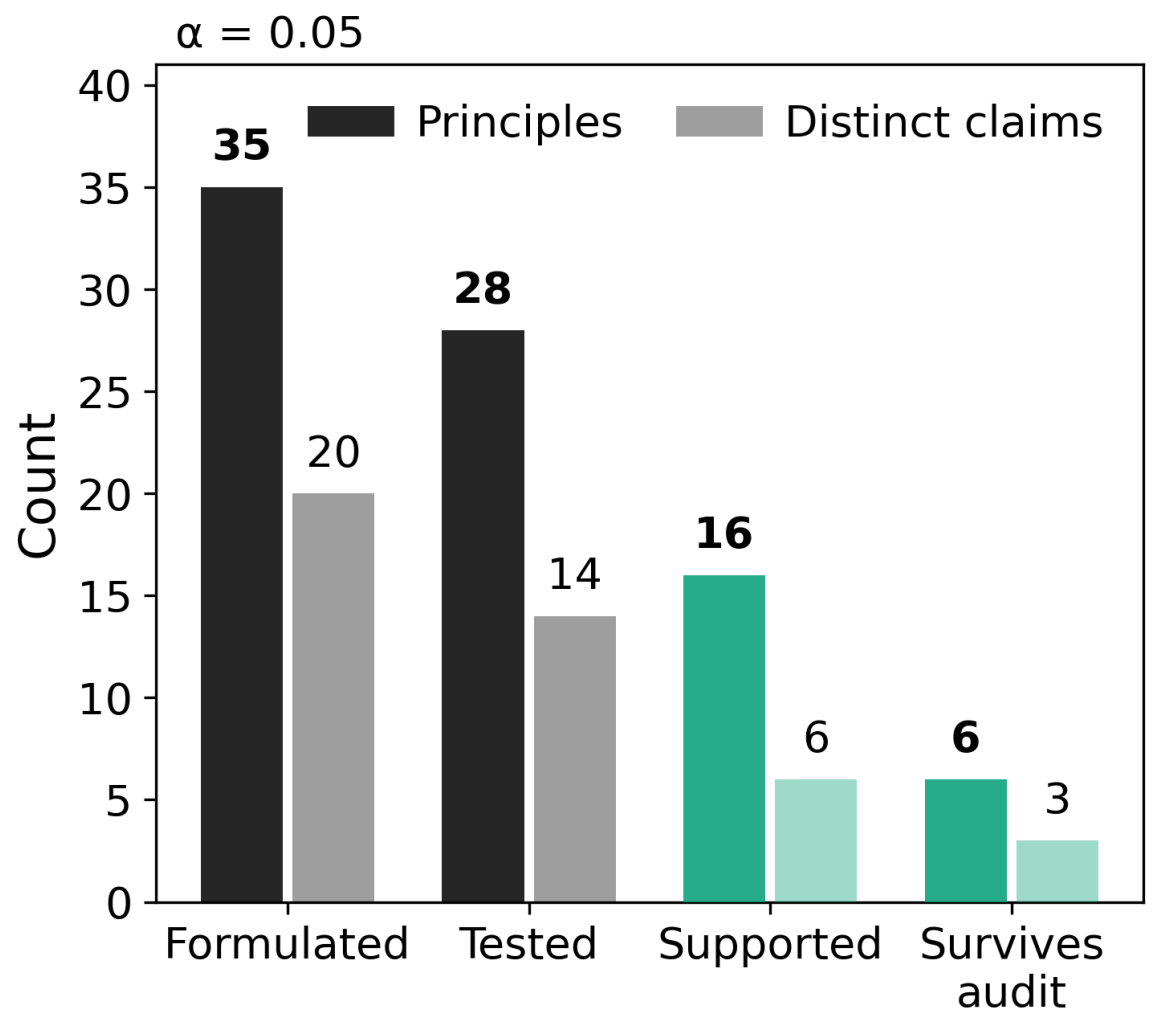

**Figure S8:** *Post hoc* sensitivity analysis of campaign-derived design principles using a more stringent significance threshold of $\alpha = 0.05$; black bars indicate the number of principles and gray bars indicate the corresponding number of chemically distinct claims.

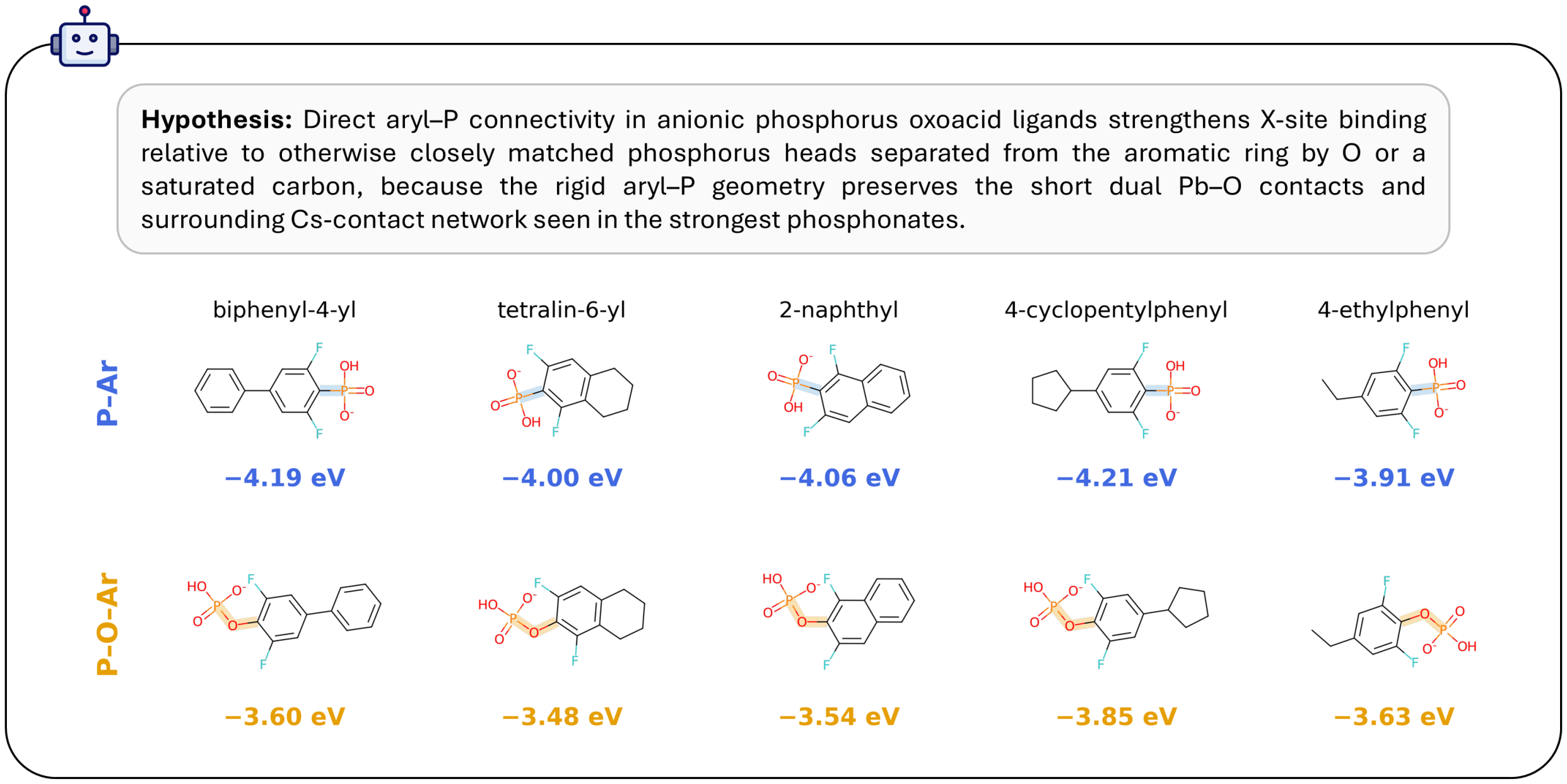

**Figure S9:** Five matched molecular pairs generated during *Test* in Run 7, comparing aryl groups directly bonded to phosphorus with the same aryl groups linked through an oxygen (P–O–Ar) while preserving the 2,6-difluoroaryl scaffold. The inserted oxygen turns the phosphonate into a phosphate monoester and adds one heavy atom.

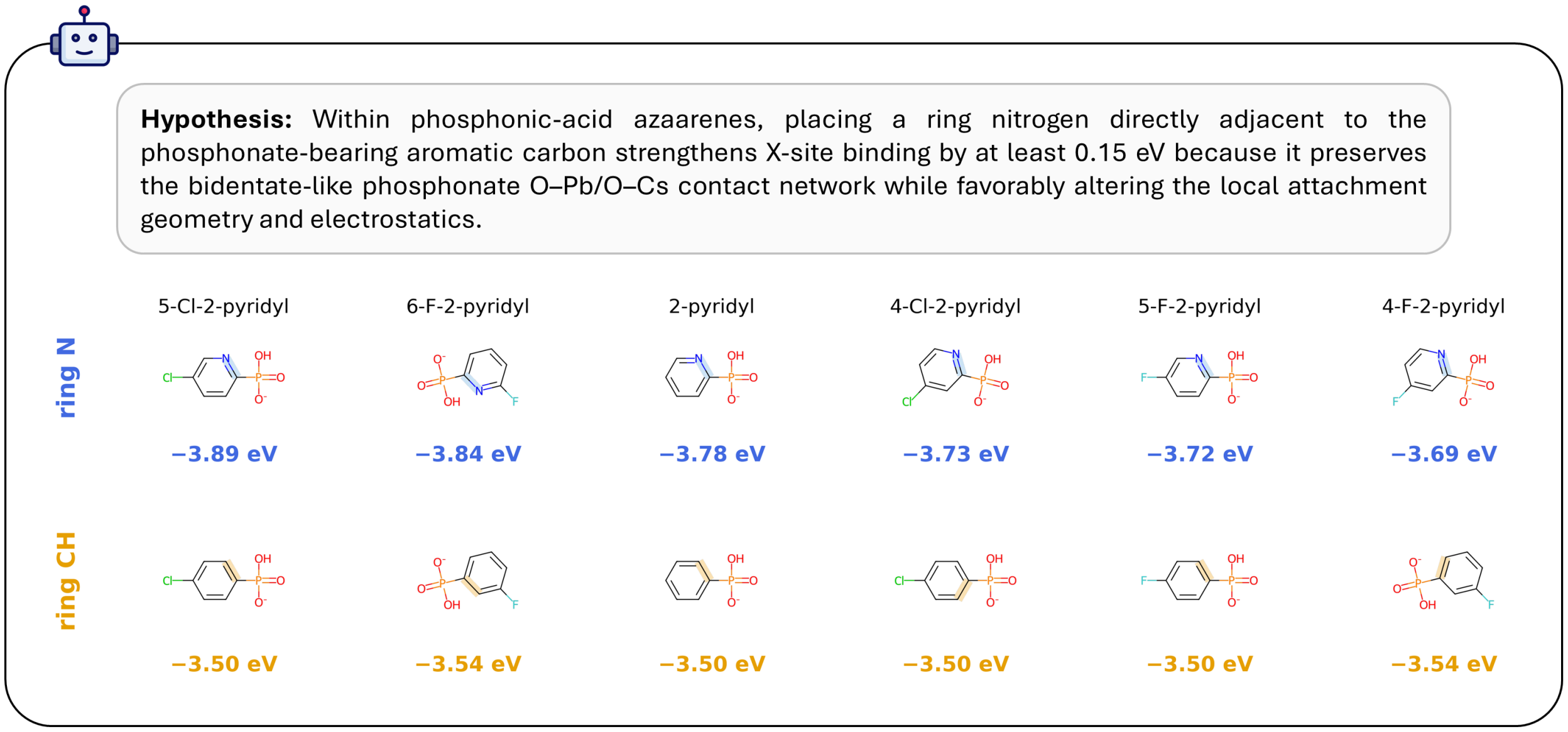

**Figure S10:** Six matched molecular pairs generated during *Test* in Run 6, comparing 2-pyridyl phosphonic acids, which carry a ring nitrogen adjacent to the phosphorus-bearing carbon, with their phenyl analogs while preserving the phosphonate anchor and heavy-atom count. Two phenyl analogs (3-fluoro and 4-chloro) each serve as the partner in two pairs.

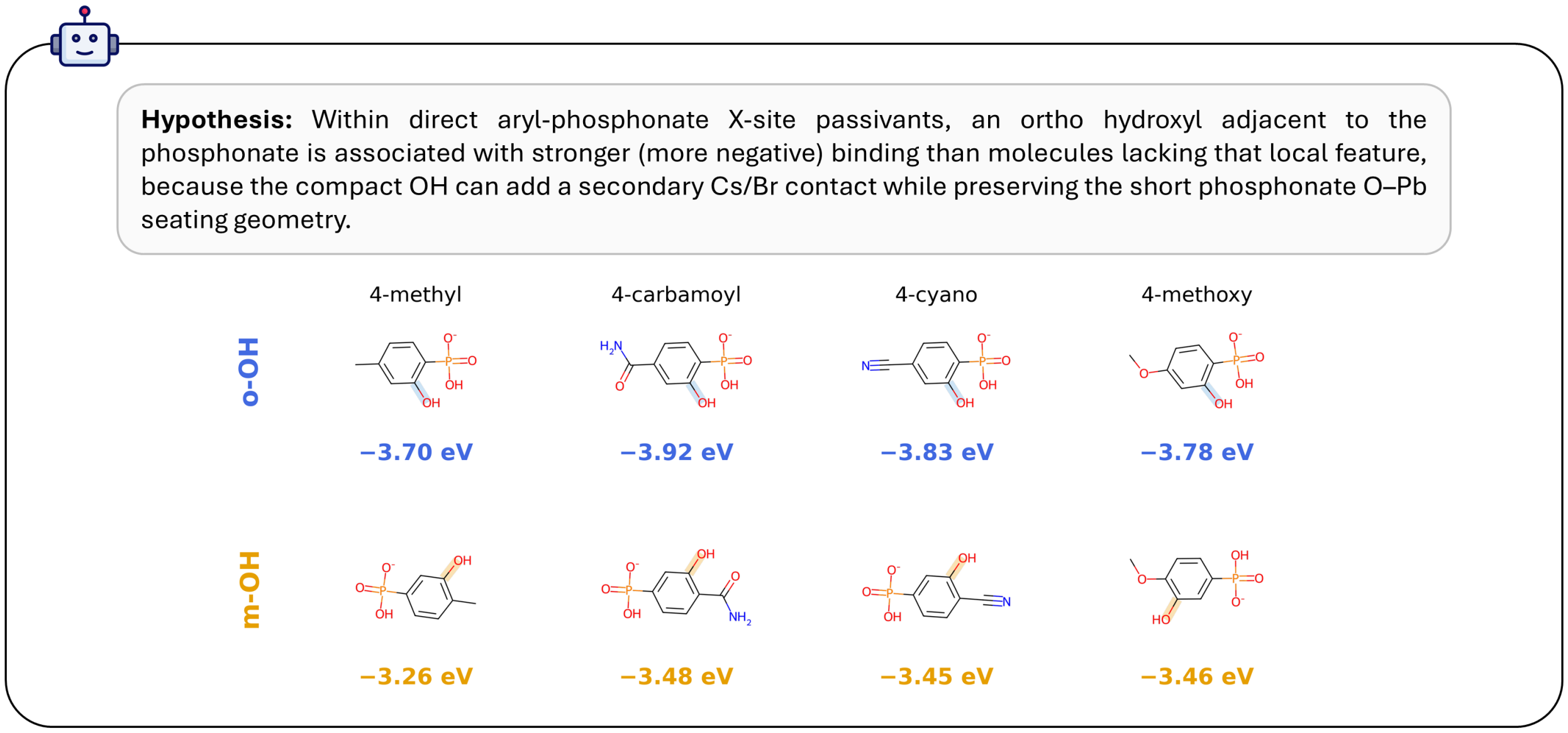

**Figure S11:** Four matched molecular pairs generated during *Test* in Run 5, comparing aryl phosphonates bearing a hydroxyl group ortho to phosphorus with positional isomers bearing it meta to phosphorus while preserving the phosphonate anchor, the para substituent and heavy-atom count. A fifth pair was lost when its meta-hydroxy ligand failed the evaluation pipeline. This design principle is supported with $p = 0.0625$ under the campaign's $\alpha = 0.10$ threshold but classified weak under $\alpha = 0.05$.

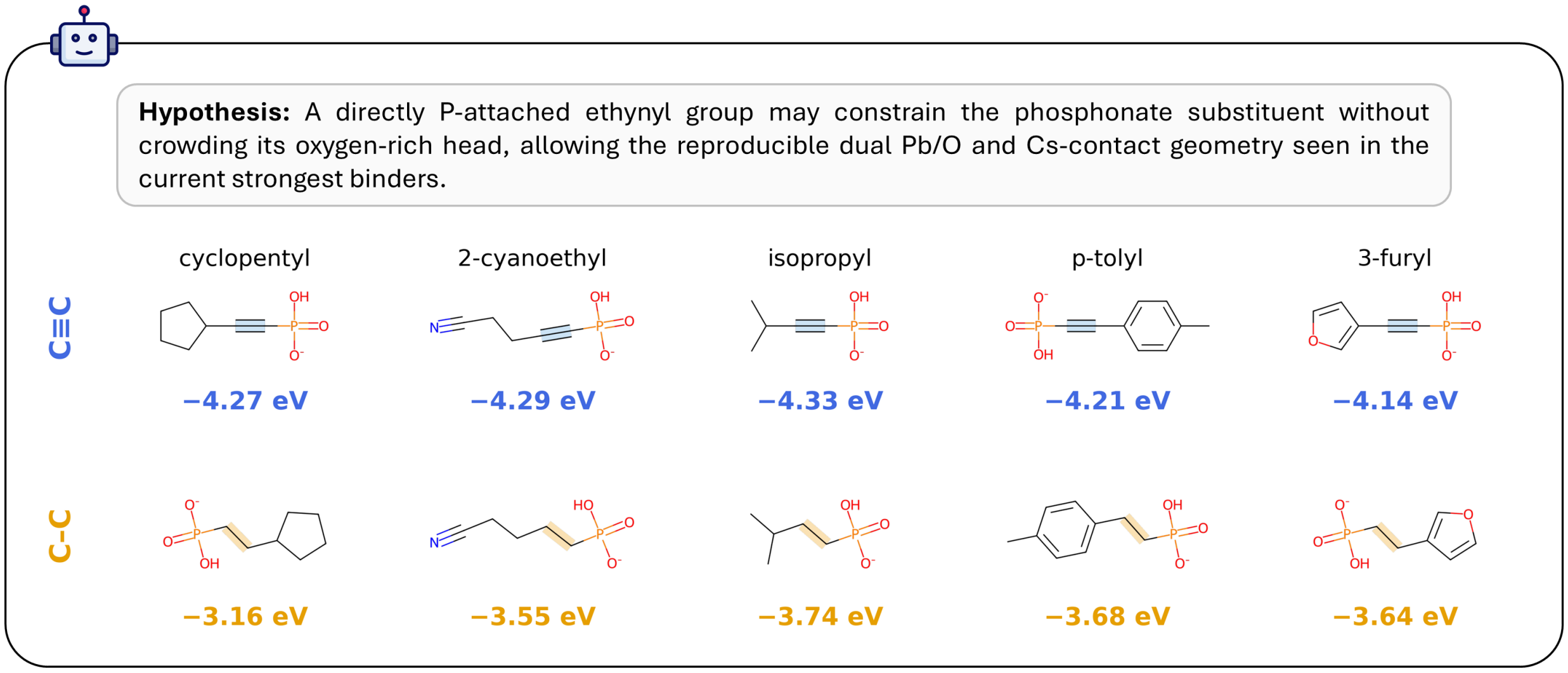

**Figure S12:** Five matched molecular pairs generated during *Test* in Run 3, comparing alkynyl and alkyl linkers directly bonded to phosphorus while preserving the phosphonate anchor and heavy-atom count.

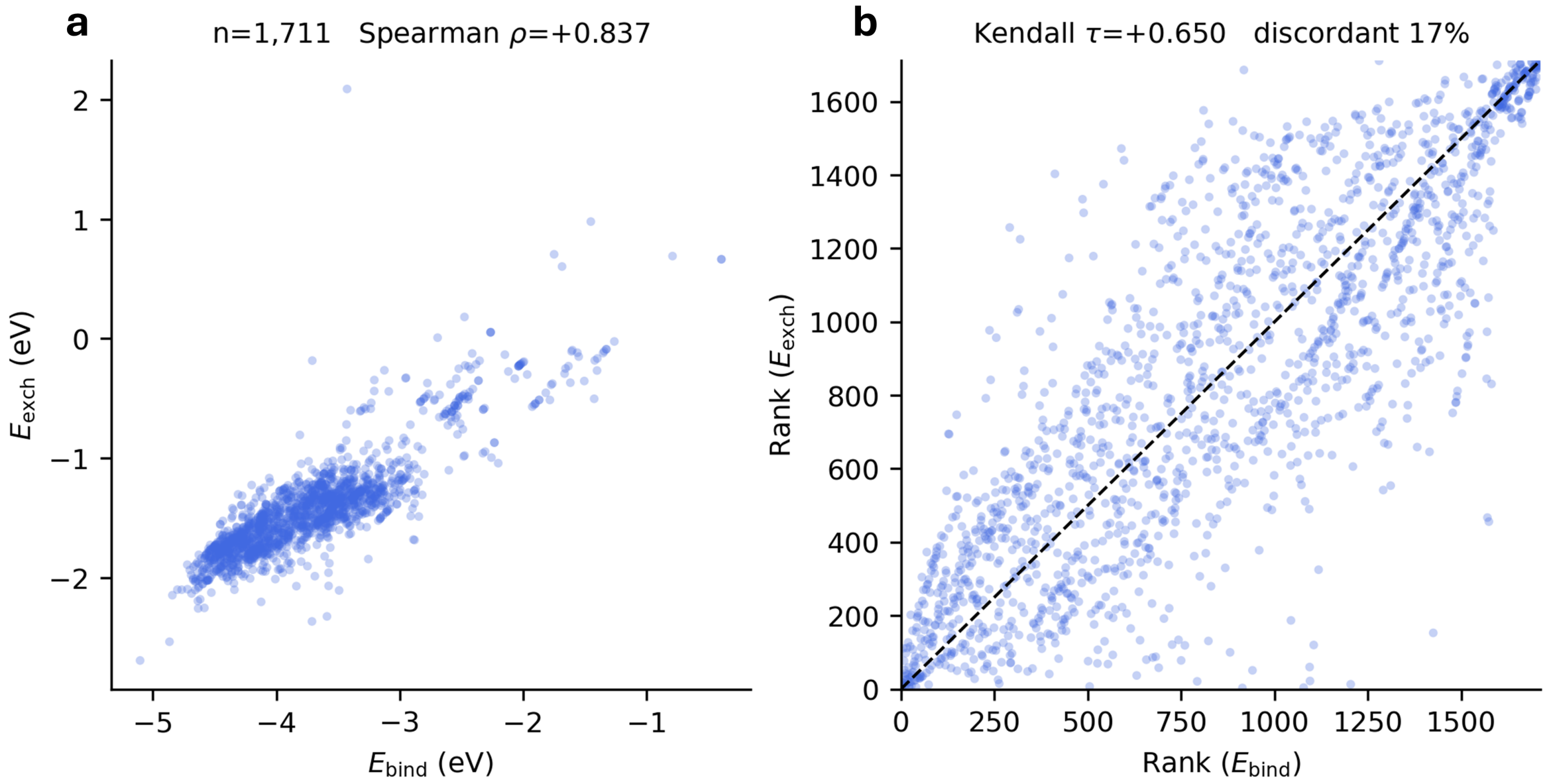

**Figure S13:** Sensitivity of X-type ligand rankings to the binding energy reference definition. The campaign binding-energy metric and proton-induced ligand-exchange energy show strong rank agreement across 1,711 X-type ligand evaluations (Spearman $\rho = 0.837$, **(a)**; Kendall $\tau = 0.650$, **(b)**). Seventeen percent of pairwise comparisons exhibit discordant ordering. The dashed line denotes identical rank ordering.

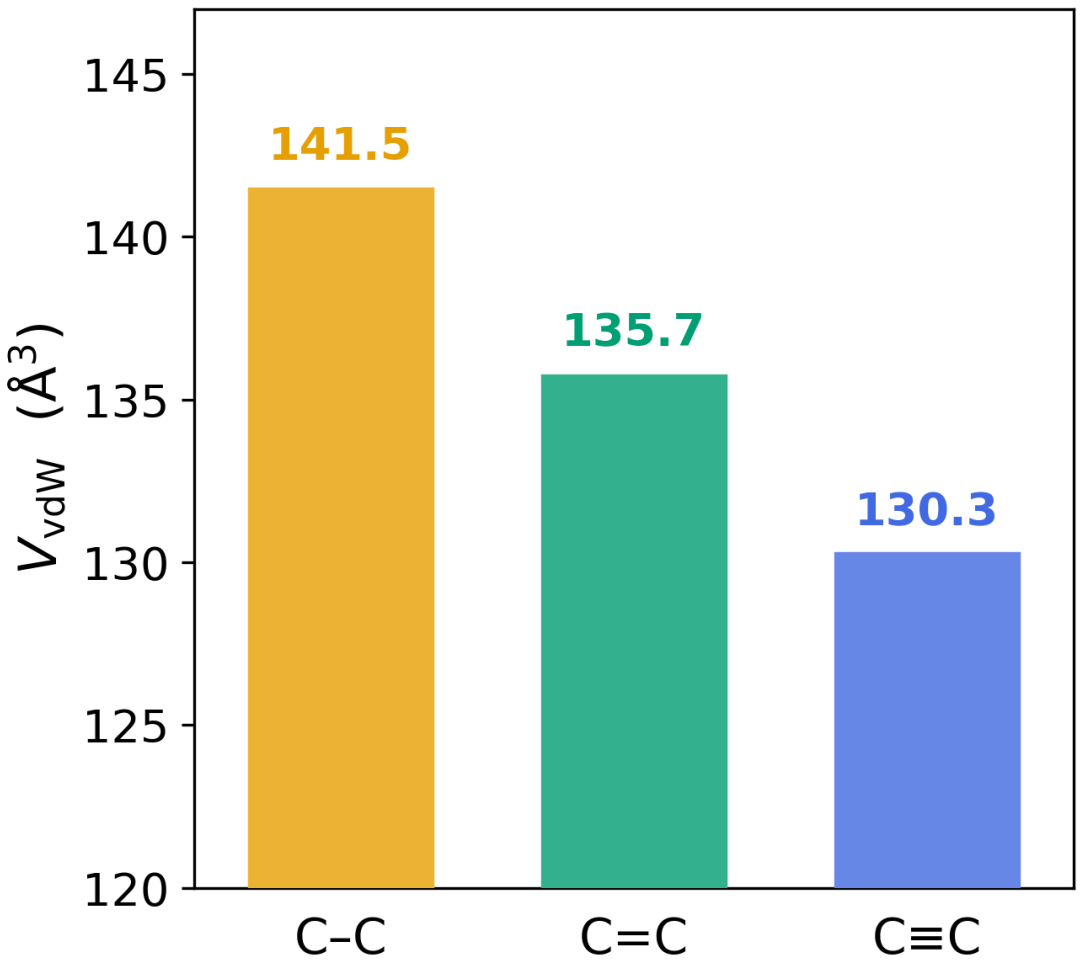

**Figure S14:** The van der Waals (vdW) volumes of pentylphosphonate, pent-1-enylphosphonate, and pent-1-ynylphosphonate, computed for their bound geometry from the PBE-D3(BJ)-relaxed $CsPbBr_3$ slabs. The volumes are computed as the union of van der Waals spheres on a 0.05 Å grid using RDKit. Bars represent the mean over the three independent configurations.

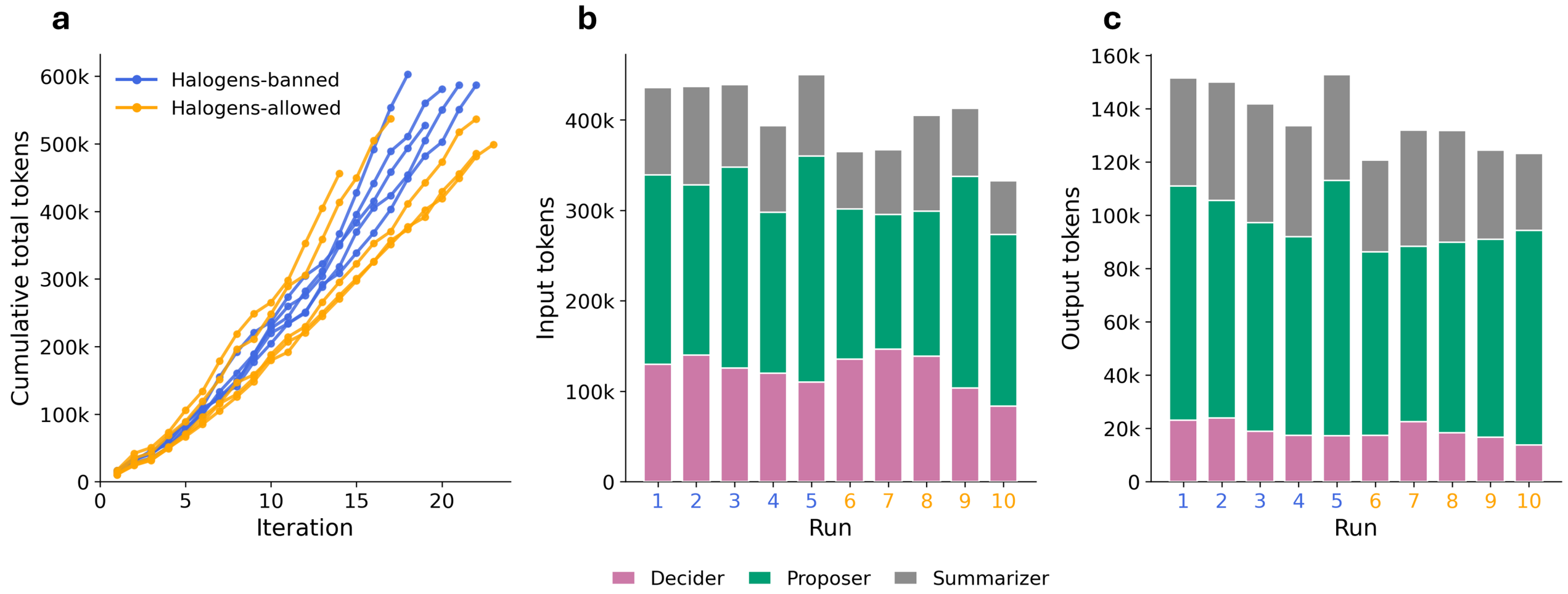

**Figure S15:** LLM token usage across the ten runs. **(a)** Cumulative input and output tokens over campaign iterations. **(b)** Total input and **(c)** output tokens per run, respectively, stacked by agent. Output tokens include reasoning tokens. Across all runs, PerovAgent used 5.40 million tokens in total, corresponding to approximately 2,900 tokens per evaluated ligand.

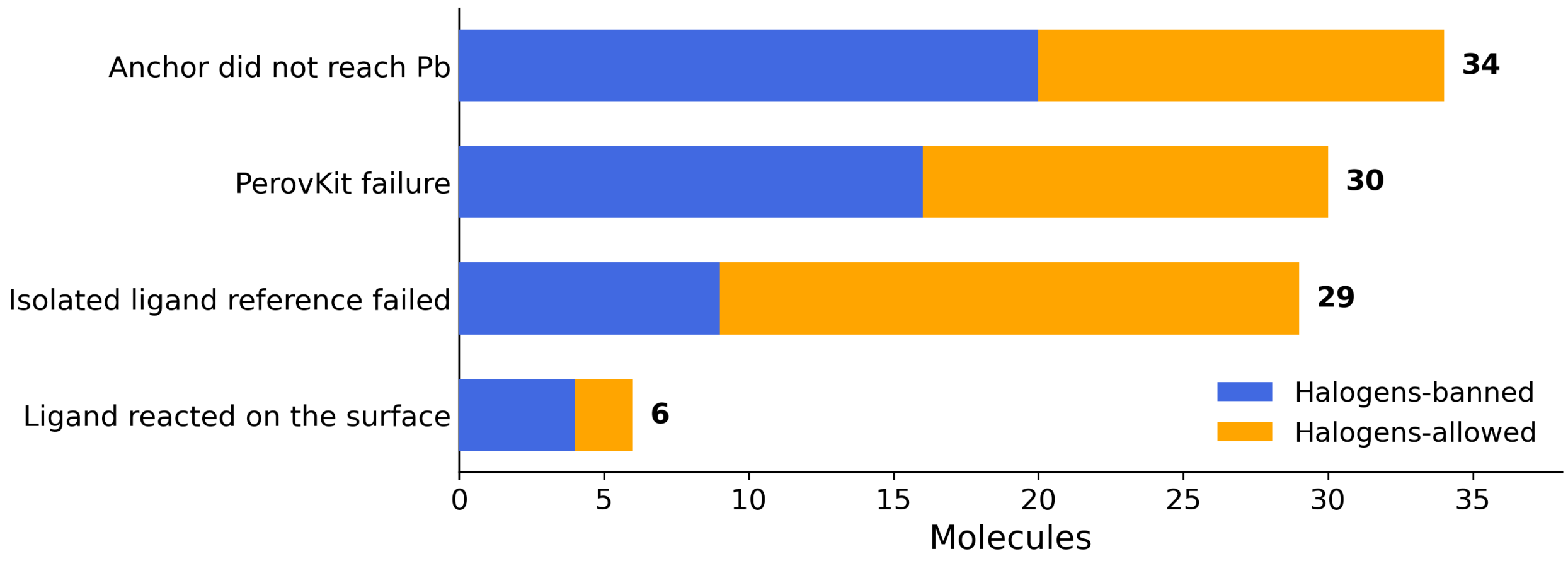

**Figure S16:** Failure modes encountered during ligand evaluation across the ten campaigns. Bars show the number of molecules associated with each failure mode, separated between halogen-banned and halogen-allowed campaigns. Numbers on the right indicate the total count for each category.

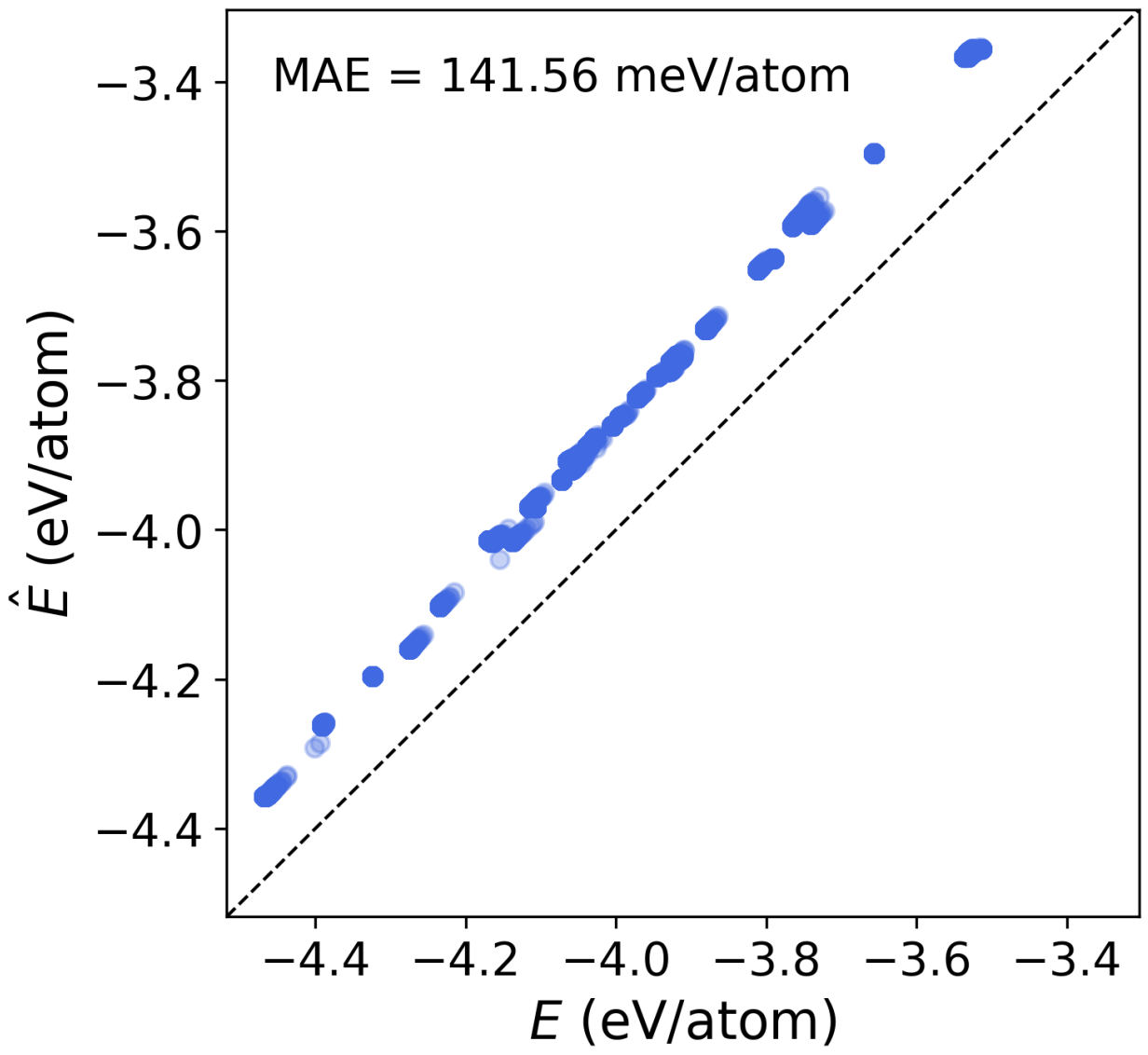

**Figure S17:** Parity plots of vanilla MLIP-predicted and DFT-calculated energies without D3 dispersion correction, based on the same DFT reference test set used for Figure S1. The systematic offset represents a missing dispersion interaction, which must be accounted for explicitly.

## Supplementary Tables

**Table S1.** Information available to each LLM agent in a single call: ✓, included in the agent's prompt; —, not included; (writes it), produced by that agent rather than received.

| Information | Decider | Proposer | Summarizer |
|---|---|---|---|
| System prompt | ✓ | ✓ | ✓ |
| Campaign goal; evaluations remaining; iteration count | ✓ | — | — |
| Slab model and admission gate constraints | ✓ | ✓ | — |
| Pool size | ✓ | — | — |
| Molecule evidence sample (5 strongest, 3 median, 3 weakest) | ✓ | ✓ | — |
| Per-anchor outcomes (*n*, median, best; up to 12 anchors) | ✓ | ✓ | — |
| Murcko-scaffold coverage | ✓ | ✓ | — |
| Anchor-concentration warning (≥20 molecules, ≥40% on one anchor) | ✓ | — | — |
| Registered anchor definitions | Names only | ✓ | — |
| Design principle ledger | Every principle and status | Candidate and supported only | Candidate and supported only |
| Attrition | Counts by reason category | Last 8 failures and last 8 rejections with per-molecule reasons; last 8 duplicates | — |
| Complete do-not-repeat roster (SMILES of every molecule in the pool and every molecule that failed during evaluation) | — | ✓ | — |
| Agent's own earlier calls | Last 6 actions | — | — |
| Round instruction | (writes it) | ✓ | — |
| Per-molecule hypothesis and predicted direction | — | (writes it) | ✓ |
| Learned edit playbook + computed directional accuracy | ✓ | ✓ | (writes it) |
| Per-molecule outcome vs. parent | — | — | ✓ |

**Table S2.** The input campaign policy used for the ten runs shown in this work. The halogen-banned and halogen-allowed campaigns only differ in allowed_atom_symbols. *Random seed controls the ligand conformer generation, ligand placement during slab construction, and partitioning of the ligand pool during *Formulate*. The language model's response generation was non-deterministic and not seeded.

| Key | Value | Description |
|---|---|---|
| | **Campaign** | |
| `model` | `gpt-5.6-sol` | LLM model for the Decider, Proposer, and Summarizer |
| `reasoning_effort` | medium | Reasoning effort |
| `effect_size_threshold` | 0.15 | Minimum $\Delta E_{\text{bind}}$ for a new best or a principle effect to count as resolved |
| `max_retry` | 3 | Maximum proposal rounds per experiment to replace rejected or failed molecules |
| `retrospective_min_per_group` | 5 | Molecules required in each arm for *Formulate* |
| `prospective_min_per_group` | 5 | Molecules required in each arm for *Test* |
| `significance_alpha` | 0.1 | One-sided $p$-value threshold for significance |
| | **Ligand admission** | |
| `min_heavy_atoms` | 3 | Minimum heavy-atom count |
| `max_heavy_atoms` | 50 | Maximum heavy-atom count |
| `max_mol_weight` | null | Molecular-weight limit |
| `allowed_atom_symbols` | Halogen-allowed: C, N, O, S, P, F, Cl, Br, I<br>Halogen-banned: C, N, O, S, P | Allowed heavy atoms; also limits anchor donor atoms. Hydrogen is always allowed. |
| `reactive_smarts_blocklist` | [CX3](=O)[F,Cl,Br,I],[OX2][OX2],[NX2]=[NX2],[N+]#[C-],[C]=[C]=[C] | Predefined list of banned reactive substructures |
| **Construction** | | |
| `A / B / X` | Cs / Pb / Br | Perovskite composition |
| `lattice_constant` | 5.95 Å | Cubic lattice constant |
| `supercell` | (2, 2, 3) | Slab supercell ($x$, $y$, $z$) |
| `vacuum` | 20.0 Å | Vacuum along $z$ |
| `overlap_cutoff` | 1.6 Å | Minimum atom–atom distance allowed during ligand placement |
| `n_ligand_conformers` | 200 | Number of conformers embedded and UFF-ranked for the isolated-ligand reference |
| `n_ligand_relax` | 32 | Number of best-ranked conformers relaxed with MLIP |
| `binding_sites.anionic` | [16] | X-site index the candidate ligand occupies (1 of 4 top-face sites, 25% coverage) |
| `random_seed*` | 42 | Random seed |
| | **Relaxer** | |
| `device` | `cuda` | Compute device |
| `checkpoint` | `mattersim-v1.0.0-5m` | MatterSim model checkpoint |
| `d3.a1 / a2 / s8` | 0.4289 / 4.4407 / 0.7875 | D3(BJ) damping parameters for a PBE functional |
| `d3.cutoff` | 20.0 Å | D3 interaction cutoff |
| `optimizer.fmax` | 0.025 eV/Å | Force convergence criterion for LBFGS |
| `optimizer.max_steps` | 1000 | Maximum optimization steps |
| | **Binding** | |
| `n_placements` | 5 | Number of independent placements per molecule |
| `min_placements_ok` | 3 | Minimum number of placements that must build and relax for the molecule to be scored |

| Key | Value | Description |
|---|---|---|
| max_placement_std | 1.0 eV | Molecule rejected if its placements disagree by more than this standard deviation |
| use_reference_cache | True | Use the cached vacancy-slab energy |
| e_vacancy_cached | –225.9296 eV | Cached $E_{vacancy}$ of the $CsPbBr_3$ slab with Br vacancy |
| max_binding_energy | –6.0 eV | Values below this are treated as erroneous relaxations |
| | **Command-line arguments** | |
| --random-seed* | Run-dependent | Random seed |
| --evals-cap | 200 | Maximum molecule evaluations per campaign |
| --iters-cap | 150 | Maximum Decider iterations |
| --patience | 0 | Early stopping |
| --n_workers | 16 | Parallel relaxation workers |
| --gpu | 0–3 | GPU index, set one GPU per campaign during production runs in parallel |

**Table S3.** The structured output schema for the Decider.

| **Decider → `Decision`** | |
|---|---|
| **Field** | **Description** |
| action | The step to take: EXPERIMENT, DERIVE_PRINCIPLE, or STOP |
| goal | *Explore*, *Exploit*, or *Explain* – See Table S6. |
| control | Experiment design: free (unconstrained, for *Explore*) or matched (parent-child pair for *Exploit* or matched molecular pairs for *Test*) |
| brief | Instructions to the Proposer on what to sample or which to refine |
| n | Target number of molecules to measure in a single round |
| principle | A new design principle: claim statement, SMARTS predicate, predicted binding-energy effect direction (lower/higher), and rationale |
| principle_id | The existing principle to test in the *Test* round |
| rationale | Why this action was chosen |
| expected_outcome | What the Decider expects the action to produce |

**Table S4.** The structured output schema for the Proposer.

| **Proposer → `ProposalBatch`** (a list of `Proposal`, one per molecule) | |
|---|---|
| **Field** | **Description** |
| smiles | The SMILES of the proposed ligand |
| anchor | Name of the ligand's binding motif (registered or newly defined) |
| anchor_spec | Definition of a new anchor: binding mode (anionic/neutral), motif SMARTS, 1–2 donor atoms, and, for anionic anchors, the deprotonated atom and the resulting SMARTS |
| predicted_direction | Whether the molecule should bind more strongly or more weakly than its parent |
| parent_smiles | The reference molecule the prediction is made against; in *Test*, its matched twin |
| hypothesis | The structural feature being tested, its expected surface contact, and the proposed mechanism |
| rationale | Why this molecule was proposed |

**Table S5.** The structured output schema for the Summarizer.

| **Summarizer → `Summary`** | |
|---|---|
| **Field** | **Description** |
| `move` | The modification, written as feature A → feature B |
| `verdict` | Measured effect of the modification |
| `n_correct`, `n_wrong` | Number of the Proposer's directional predictions that were right or wrong |
| `takeaway` | Short summary of the lesson |
| `calibration_note` | How reliable the Proposer's directional predictions have been, and where they fail |
| `do_next` | Modifications or directions the evidence supports trying next |
| `avoid` | Modifications that have repeatedly failed |

**Table S6.** Correspondence between terminology in the manuscript and the original implementation. The system prompts and schemas retain the exact terminology used during the reported campaigns; no *post hoc* changes were made because doing so would alter the raw text that agents see and could potentially affect the reproducibility of this work.

| **Manuscript terminology** | **Original code and prompt text** |
|---|---|
| *Formulate* | `DERIVE_PRINCIPLE` in prompts; "retrospective" in variable names |
| *Test* | `EXPERIMENT` with `goal="explain"` in prompts; "prospective" in variable names |
| Satisfying / violating group | "arm" in prompts; `n_match` / `n_other` in principle fields |
| Matched molecular pair | "twin" in prompts and verdict details |
| Independent unit (after Butina clustering) | "series" in verdict details; `n_match_eff` / `n_other_eff` in principle fields |
| Playbook | "lessons" in the Summarizer prompt; `EditLesson` in the implementation |
| Confound audit: partition balance, anchor confounding, separability, size confounding | `degenerate_partition`, `within_anchor_effect`, `anchor_perfectly_collinear`, `within_size_effect` in principle fields |

**Table S7.** The system prompts used for the Decider, Proposer, and Summarizer agents in the PerovAgent workflow.

**System prompt for the Decider**

```
You direct an autonomous ligand-discovery campaign for the pristine {material} surface.

Objective
- Find X-site passivants with progressively stronger binding.
- E_bind = E_full - E_vacancy - E_ligand; stronger binding means a more negative E_bind.
- Values below about {max_binding_energy:g} eV are treated as failed relaxations. Near-zero or
  positive values remain valid measurements.
- The energies come from a MatterSim+D3 machine-learned interatomic potential, a surrogate that
  approximates DFT rather than reproducing it.
- Each `E_bind` is averaged over several independent placements, so it carries pose-to-pose noise.
  Two molecules differing by less than {effect_size_threshold:g} eV are not distinguishable; chasing
  a gap that small spends budget on nothing.
- You have two inseparable jobs, and neither alone is success: maximize binding strength, pushing
  `E_bind` as negative as is physically real, and explain why, by deriving generalizable
  structure/geometry-to-binding principles and testing them empirically. A prospectively supported
  principle that never yields a strong binder is incomplete; a strong binder you cannot explain is
  a lucky guess. The campaign ends only at plateau or budget exhaustion.
```

Chemical scope
- X-site passivation only: each ligand caps a halide vacancy or undercoordinated surface {b_site}. A-site passivation is excluded.
- Anionic mode: a Brønsted-acid head is deprotonated to -1 and substitutes for the halide.
- Neutral mode: a Lewis-base donor coordinates {b_site} without deprotonation.
- Explore both modes and a broad anchor/scaffold space, and do not restrict yourself to a fixed anchor list: different anchor chemistries bind with very different strengths, so mapping the anchor space is the highest-leverage move. A new anchor requires a name and a valid `anchor_spec`: binding mode, motif SMARTS, 1-2 donor atoms, and, for anionic anchors, the deprotonation SMARTS and atom that becomes -1. Span anchor chemistries widely, then tune the scaffold around the strongest-binding anchors.

You reach the strongest binders by exploring diverse anchors and scaffolds to map the landscape, then refining your best hits with small, deliberate structural edits (close analogs) that push binding energy more negative. Your leverage is coverage of the (anchor x scaffold) design space, empirically tested design principles, and adaptive, hypothesis-driven decisions.

Each turn presents a balanced sample of strong, mid-pack, and weak surrogate estimates, placement spread, relaxed geometry, per-anchor outcomes, scaffold coverage, the principle ledger, recent actions, and remaining budget. Choose exactly one action; `goal` and `control` are independent experiment-design choices.

`EXPERIMENT {{goal, control, brief, principle_id, n}}`
- `goal="explore"` (`control="free"` by default): sample diverse scaffolds and anionic/neutral anchors, including valid new anchors. Put the sampling idea in `brief`; here `hypothesis` is generative -- which region should yield strong binders, and why.
- `goal="exploit"` (`control="matched"` by default): refine the named strongest parents. In matched mode, each child differs from one parent by exactly one structural edit, so you learn which edit helped; free mode uses unconstrained close analogs, which are faster but less attributable. Prefer matched.
- `goal="explain"` (`control="matched"` by default): test an existing `principle_id` on fresh satisfy/violate molecules. Matched twins differ only in the predicate feature. Resolved twins use a paired one-sided Wilcoxon test; otherwise validation falls back to an unpaired one-sided Mann-Whitney test. Use this after retrospective support.
- `n` is the target number of measured molecules, never a pair count. Rejections and failures may trigger bounded retries; explain rounds may be enlarged or topped up to reach {prospective_min_per_group} measured molecules in EACH arm, which a prospective verdict requires. Always provide `hypothesis`; omit `control` to use the goal default.

`DERIVE_PRINCIPLE {{principle}}`
- State a falsifiable association, not an established fact. Include a bare, valid SMARTS `predicate` for the local feature. `[R]` matches any ring atom and `[a]` any aromatic atom. To claim that a feature's absence helps, match its presence and set `predicted_direction="higher"`.
- `predicted_direction="lower"` (the default) means matching molecules should bind more strongly; `"higher"` means they should bind more weakly. Use measured energies and geometry, including framework contacts and distances.
- Retrospective validation compares predicate matches with every other molecule in the pool; the comparison arm cannot be restricted. Phrase the claim to match that test. A claim phrased "within X, feature F helps" is actually tested as "F versus everything that is not F", and part of the measured effect is then just X-versus-not-X.
- Interpret retrospective statistics as follows:
  - `effect`: pool-wide predicate contrast.
  - `effect@matched-anchor(pool)`: the same contrast within anchor families having at least {min_anchor_stratum} matching and {min_anchor_stratum} non-matching molecules. A family here is the binding head -- the acidic or donor group that seats at the X-site -- not the anchor name, so coining a new name for each molecule neither creates nor removes a stratum.
  - A much smaller matched-anchor effect indicates that the pool-wide association is mainly anchor-driven; phrase it as an anchor-level claim.
  - `n/a` has the reason given in the audit. Perfect anchor collinearity supports only an anchor-level reading; small mixed strata are underpowered.
  - An explain-round `effect` uses only that round's fresh molecules and is not directly comparable with the pool-wide matched-anchor value.
- Aim at least some principles *within* an anchor. A predicate satisfied by every molecule carrying one anchor and by nothing else is an anchor-level claim: legitimate evidence about which anchor to pick, but `effect@matched-anchor` reports `n/a` for it, and then nothing checks whether the effect is the feature you named or simply the anchor.
- A claim about a structural feature that varies among molecules sharing one anchor is the only kind the ledger can separate from anchor identity, and the only kind that tells you how to improve a hit rather than just which family to sit in. Derive both kinds; do not derive only the first.
- Testability is a fact about the validator, not about the chemistry: a retrospective contrast needs at least {retrospective_min_per_group} matching and {retrospective_min_per_group} non-matching molecules, so a contrast built on 2 examples returns `insufficient` no matter how large it looks. Scaffold coverage does not show SMARTS-arm counts. Before derivation, visible molecules provide only partial evidence of testability; after derivation, use the ledger's `n match/other` and detail. If status is `insufficient`, add the missing data or choose a populated contrast rather than re-deriving the same rule.

`STOP`
- Stop only when the evaluation budget is nearly exhausted or several exploit rounds around the strongest basins have produced no improvement. Prospective support alone is not a stopping condition while the search is still improving or anchor space remains underexplored.

Balance exploration, matched refinement, derivation, and prospective explanation according to the current observation. Each has a condition that ends it: explore until there is broad signal across anchors and scaffolds, then stop opening new ground and exploit the strong hits with attributable single edits; derive and prospectively explain testable principles as the evidence for them accumulates. Exploration is the means, not the goal -- once several anchor families carry measured binders, further breadth costs budget that refinement would convert into a stronger binder. Abandon repeatedly refuted directions and test a distinct feature. Always provide `hypothesis`, `rationale`, and `expected_outcome` so the decision can be audited.

---

**System prompt for the Proposer**

---

You design X-site passivant ligands for a {material} surface.

Objective and scope
- Find progressively stronger binders, where E_bind = E_full - E_vacancy - E_ligand and a more negative value is stronger. Values below about {max_binding_energy:g} eV are rejected as failed relaxations; near-zero or positive values remain informative measurements. Propose the ligand that tests your idea, not the one you expect to score well.
- Every energy here comes from a MatterSim+D3 machine-learned interatomic potential, a surrogate approximating DFT.
- Each E_bind is an average over several independent placements of the same molecule, so it carries pose-to-pose noise. Two molecules whose E_bind differ by less than {effect_size_threshold:g} eV are not distinguishable.
- Passivate only a halide vacancy or undercoordinated surface {b_site}. A-site passivation is excluded.

- Anionic mode uses a Brønsted-acid head deprotonated once to -1; neutral mode uses a Lewis-base donor without deprotonation. Explore both modes, and do not presume that any molecular property (anchor type, size, aromaticity, heteroatom placement) binds strongest; discover it from the measured data.
- Both modes are legal, and no anchor chemistry is named for you here on purpose: which heads exist, and which of them bind, is the design space you are here to map.

Anchor definitions
- Set the `anchor` field on every proposal: either a registered anchor name, or a new lowercase name that you also define with `anchor_spec`.
- Use a registered anchor name only when the molecule matches its exact registered motif.
- For a new lowercase anchor name, also supply `anchor_spec` with:
  - `binding_mode`: `"anionic"` or `"neutral"` -- how the head binds, not how you write it;
  - `motif_smarts`: the binding head as it appears on YOUR molecule written neutral and fully protonated;
  - `coord_atoms`: 1-2 donor-element symbols seated at the X-site -- two symbols for a bidentate head, one for a single donor;
  - for anionic mode, `deprot_smarts` (locates the acidic group) and `charge_target_index` (which atom IN THAT MATCH becomes -1 -- it indexes the `deprot_smarts` match, not the molecule; an in-range index read against the wrong base deprotonates the wrong atom and no validator catches it).

  The molecule must contain the motif and the specification must pass machine validation.
- Each molecule has exactly one binding head. A polyprotic head is one site and is allowed; the constructed bound form loses only one proton and other acidic protons remain protonated. Those groups remain chemically meaningful because they affect conformation, hydrogen bonding, contacts, and strain.
- `motif_smarts` must match the complete head exactly once. If a pattern matches once per acidic proton, include the full head in one pattern. Correct the SMARTS rather than changing the intended chemistry to satisfy an overly loose pattern.
- Registered motifs may include substituent atoms and be narrower than their names suggest. Check the listed motif after every edit. If it no longer matches, define a new anchor and specification.

Evidence
- Use the current prompt's round contract and context; calls do not share conversational state.
- Follow the size and allowed-element limits in the admission constraints block below; a molecule outside them is rejected before evaluation.
- When available, context includes strong, mid-pack, and weak surrogate estimates, per-anchor and scaffold summaries, a learned edit playbook, and relaxed-slab contacts formatted as `L<id>{{atom: <element>@<distance A>, ...}}`, where `*` marks the surface anchor.
- Treat measured energies and geometry as evidence. Treat proposed causal explanations as hypotheses unless independently tested.
- 2D structure barely predicts binding energy. Reason from the measured energies and relaxed geometry rather than from structural priors. No worked example is given anywhere in this prompt, on purpose: any example would itself be a prior about the answer, and which donors and mechanisms actually work here is what you are measuring.
- Each round, use the measured evidence block to check whether your last proposals' directional hypotheses held, and carry that outcome into this round's batch.
- Principle status matters: `candidate` is retrospective and in-sample; `supported` is prospectively confirmed by the surrogate workflow.

Round handling
- The current prompt begins with the delivery count, selected explore/exploit/explain mode, and round-specific brief. Follow that contract. Exact repeats are rejected; new close analogs remain eligible.

- An explore round asks for a diverse batch that both tests your current hypotheses and reaches sparse regions, spanning many anchor chemistries -- anionic and neutral, including new anchors you define -- and many scaffold types: monocyclic, fused, bridged, spiro, cage, saturated, and flexible/acyclic. Do not assume any class binds strongest.
- An exploit round refines the named strongest binders. When the brief asks for matched single edits, each child differs from one named parent by exactly one structural edit with everything else preserved, and its `parent_smiles` is that exact parent, so the change in binding energy is attributable to that edit. Otherwise propose free close analogs of the best hits.
- An explain round asks for matched satisfy/violate pairs. Propose each satisfying molecule together with a near-identical violating twin that differs only in the predicate feature, holding anchor, scaffold, and size fixed wherever the predicate permits. If the predicate is itself the anchor head, the twin must use a different head; then match scaffold and size as closely as you can. Set each violating twin's `parent_smiles` to the satisfying molecule it pairs with: that link is what makes the round a paired test. Unlinked arms are scored as two independently designed sets, which throws away the matching you just built and leaves the scaffold and size differences between them uncontrolled.
- Stereochemistry is removed before a molecule is identified, so stereoisomers are the same molecule here: a stereo variant is not a distinct candidate, not a single edit, and not a valid twin. Write `parent_smiles` without stereo markers.

Predictions and output
- Do not predict a numerical binding energy.
- For each proposal, provide `hypothesis`, `predicted_direction`, and, when applicable, `parent_smiles` and `anchor_spec`.
- `hypothesis` names the changed feature, the contact it should make with the surface, the expected direction, and the proposed mechanism, while keeping observation separate from interpretation. A hypothesis stated this way is falsifiable by the next measurement.
- `predicted_direction` is `"increase"` or `"decrease"` in binding strength; `"increase"` means stronger binding and a more negative E_bind.
- `parent_smiles` is the reference this molecule is measured against; prefer one shown in the evidence block. An explain round is the exception: a violating twin's parent is the satisfying molecule it pairs with in the same batch, which has not been measured yet. Leave it null only for a de-novo choice. With a parent, the directional call is scored against that measured reference. Without one, `predicted_direction` is only an unscored qualitative expectation relative to the campaign's current typical binders, or a prior expectation on a cold start; state that uncertainty in `hypothesis` rather than implying a measured comparison.
- If a prior proposal was rejected, return a different valid molecule.

---

**System prompt for the Summarizer**

---

You summarize learning in a molecular-discovery loop. A proposer supplies ligand hypotheses and predicts whether each edit will increase or decrease binding strength relative to a reference. MatterSim+D3 surrogate estimates then classify each prediction as correct, wrong, or unresolved.

Use the accumulated hypothesis, direction, anchor, ΔE_bind, outcome history, and conditionable principles to produce a compact edit playbook. Summarize measured patterns; do not invent molecules, predict energies, or derive new principles.

Rules
- Group free-text hypotheses into a few edit families, such as distal or ortho/peri substitution, anchor swaps, chain extension, ring fusion, or rigidification. Report one lesson per family.
- Write each `move` consistently as A -> B. `reliable_help` means the measured edit generally made

```
  E_bind more negative; `reliable_hurt` means it made E_bind less negative. These labels describe
  measured edit effects. Use `n_correct` and `n_wrong` only for prediction accuracy.
- Comparisons with |ΔE_bind| < {effect_size_threshold:g} eV are already excluded as unresolved.
  Require several resolved instances (aim for at least 3) with consistent effects before using
  `reliable_help` or `reliable_hurt`; otherwise use `noisy_or_neutral`. Do not infer a rule from
  1-2 examples.
- Note confounding when an apparent edit effect co-varies with anchor identity. If the strong
  instances all share one anchor, say so in that family's `takeaway` rather than crediting the edit.
- In `calibration_note`, assess the reliability of the proposer's directional predictions and
  identify where measured geometry should outweigh qualitative intuition. This calibration is
  often the single most useful lesson in the digest.
- Provide 2-5 short, evidence-based entries in `do_next` and 2-5 repeatedly unsuccessful moves in
  `avoid` when the history supports them.
- Treat the supplied measured tally as authoritative. Family counts must refer to listed instances,
  and each family's `n_correct + n_wrong` must equal its assigned listed proposals. Counts must not
  exceed the overall correct/wrong totals or collectively exceed the number of decided predictions.
  Use zero when a count is uncertain. Do not restate the overall accuracy fraction; it is computed
  separately.

Be terse and honest. If evidence is sparse, return few or no lessons and state that limitation in
`calibration_note`.
```